\documentclass[prd,twocolumn,superscriptaddress,amsfonts,amssymb,amsmath,showpacs]{revtex4-2}
\usepackage{bm}
\usepackage{amsfonts}
\usepackage{latexsym}
\usepackage[latin1]{inputenc}
\usepackage{graphicx}
\usepackage{amsmath}
\usepackage{palatino}
\usepackage{mathpazo}
\usepackage[british]{babel}
\usepackage{hhline}
\usepackage{multirow}
\usepackage{textcomp}
\usepackage{float}
\usepackage{booktabs}
\usepackage{dcolumn}
\usepackage{ragged2e}
\usepackage{hyperref}
\hypersetup{colorlinks,citecolor=blue}
\hypersetup{colorlinks=true,linkcolor=red,filecolor=magenta,    urlcolor=cyan}
\usepackage{amsmath}
\usepackage{xcolor}
\usepackage{orcidlink}
\usepackage{epsfig}
\usepackage{caption}
\usepackage{subcaption}
\usepackage{commath}
\def\jnl@style{\it}
\def\aaref@jnl#1{{\jnl@style#1}}

\def\aaref@jnl#1{{\jnl@style#1}}

\def\aj{\aaref@jnl{AJ}}                   
\def\apj{\aaref@jnl{ApJ}}                 
\def\apjl{\aaref@jnl{ApJ}}                
\def\apjs{\aaref@jnl{ApJS}}               
\def\apss{\aaref@jnl{Ap\&SS}}             
\def\aap{\aaref@jnl{A\&A}}                
\def\aapr{\aaref@jnl{A\&A~Rev.}}          
\def\aaps{\aaref@jnl{A\&AS}}              
\def\mnras{\aaref@jnl{Mon.~Not.~Roy.~Astron.~Soc.}}             
\def\prd{\aaref@jnl{Phys.~Rev.~D}}        
\def\prc{\aaref@jnl{Phys.~Rev.~C}}  
\def\prl{\aaref@jnl{Phys.~Rev.~Lett.}}    
\def\qjras{\aaref@jnl{QJRAS}}             
\def\skytel{\aaref@jnl{S\&T}}             
\def\ssr{\aaref@jnl{Space~Sci.~Rev.}}     
\def\zap{\aaref@jnl{ZAp}}                 
\def\nat{\aaref@jnl{Nature}}              
\def\aplett{\aaref@jnl{Astrophys.~Lett.}} 
\def\apspr{\aaref@jnl{Astrophys.~Space~Phys.~Res.}} 
\def\physrep{\aaref@jnl{Phys.~Rep.}}      
\def\physscr{\aaref@jnl{Phys.~Scr}}       
\def\commat{\aaref@jnl{Comm.~Math.~Phys.}}              
\def\science{\aaref@jnl{Science}}               
\def\cqg{\aaref@jnl{Classical Quant.~Grav.}}            
\def\jpcs{\aaref@jnl{JPCS}}                                     
\def\ijmpd{\aaref@jnl{Int.~J.~Mod.~Phys.~D}}                    
\def\grg{\aaref@jnl{Gen.~Relat.~Gravit.}}               
\def\rpp{\aaref@jnl{Rep.~Prog.~Phys.}}          
\def\npa{\aaref@jnl{Nucl.~Phys.~A}}        
\def\lrr{\aaref@jnl{Living Rev.~Rel.}}                   
\def\jcap{\aaref@jnl{J.~Cosmology Astropart.~Phys.}}    
\def\rmp{\aaref@jnl{Rev.~Mod.~Phys.}}   
\def\epjc{\aaref@jnl{Eur.~Phys.~J.~C}} 
\def\plb{\aaref@jnl{~Phy.~Lett.~B}} 
\def\mpla{\aaref@jnl{Mod.~Phy.~Lett.~A}} 
\def\arxiv{\aaref@jnl{arxiv.org}}

\allowdisplaybreaks[1]

\begin{document}
\color{black}       
\title{\bf Analyzing the geometrical and dynamical parameters of modified Teleparallel-Gauss-Bonnet model}

\author{Santosh V Lohakare \orcidlink{0000-0001-5934-3428}}
\email{lohakaresv@gmail.com}
\affiliation{Department of Mathematics,
Birla Institute of Technology and Science-Pilani, Hyderabad Campus,
Hyderabad-500078, India.}

\author{B. Mishra \orcidlink{0000-0001-5527-3565}}
\email{bivu@hyderabad.bits-pilani.ac.in}
\affiliation{Department of Mathematics,
Birla Institute of Technology and Science-Pilani, Hyderabad Campus,
Hyderabad-500078, India.}

\author{S. K. Maurya \orcidlink{0000-0003-4089-3651}}
\email{sunil@unizwa.edu.om}
\affiliation{Department of Mathematical and Physical Sciences,
College of Arts and Sciences, University of Nizwa, Nizwa, Sultanate of Oman}

\author{Ksh. Newton Singh \orcidlink{0000-0001-9778-4101}} 
\email{ntnphy@gmail.com} 
\affiliation{Department of Physics, National Defence Academy, Khadakwasla, Pune 411023, India}

\begin{abstract}
\textbf{Abstract}: To recreate the cosmological models, we employed the parametrization approach in modified teleparallel Gauss-Bonnet gravity. It has been interesting to apply the parametrization approach to investigate cosmological models. The real benefit of using this method is that the observational data may be incorporated to examine the cosmological models. Several cosmological parameters were examined, such as the Hubble parameter $(H)$, the deceleration parameter $(q)$, and the equation of state (EoS) parameter $(\omega)$. The results obtained are consistent with recent cosmological findings in the conventional scenario. A transition scenario from a decelerating stage to an accelerating stage of cosmic evolution has been observed. The EoS parameter is also in the quintessence phase, which drives the accelerating expansion of the Universe. Also, we look at the violation of strong energy conditions, which has become inevitable in the context of modified gravitational theory. Finally, we have performed the $Om(z)$ diagnostic and also obtained the age of the Universe by using the data from the cosmological observations.  
\end{abstract}

\maketitle

\section{Introduction} \label{SEC I}
The suggestions from cosmological observations and theoretical arguments convey that the early phase of the Universe has undergone an inflationary stage and in the late phase experienced the accelerated phase \cite{Copeland06, Cai10}. One can introduce the concepts of inflation and dark energy (DE) to achieve this. Another way is to modify the gravitational sector \cite{Capozziello11}, which resulted in modified gravity theories. Usually, the curvature-based Einstein-Hilbert action is extended in most modified or extended gravitational theories. Another approach to modifying the gravitational action is to extend  the equivalent torsion formulation, known as the teleparallel equivalence of general relativity (TEGR) \cite{Einstein28a, Einstein28b, Arcos04, Maluf13, Aldrovandi13}. In this class of modified gravity, the curvature less Weitzenb$\ddot{o}$ck connection is used in place of the torsion less Levi-Civita connection, i.e., in place of  curvature, attributing gravity to torsion.  As in general relativity (GR), the curvature scalar is obtained by contacting the curvature tensor, in this kind of modified gravity the torsion scalar $T$ can be obtained from the contraction of the torsion tensor. The $f(T)$ extensions of TEGR can be constructed \cite{Ferraro07, Bengochea09, Linder10} in the same approach as the $f(R)$ extension of GR has been constructed \cite{Felice10, Nojiri11}. To note though, TEGR coincides with GR; several important studies on the astrophysical and cosmological aspects of $f(T)$ gravity are done in the literature \cite{Duchaniya22, Capozziello11a, Miao11, Bamba11, Cai16, Myrzakulov11, Baojiu11, Tamanini12, Anagnostopoulos19, DeBenedictis22, Nair22}.

The cosmological observations \cite{Riess98, Perlmutter99} revealed the accelerated expansion of the Universe at least at the late phase of the cosmic evolution. Using the explosions of white dwarfs, astronomers measured the expansion of the Universe which leads to the discovery of an unknown form of energy, called dark energy. The cosmic microwave background (CMB) further revealed that DE contributes $68\%$ to the total energy content of the Universe. The cosmological constant $\Lambda$ is the most straightforward dark energy candidate with the EoS parameter, $\omega=-1$. The $\Lambda$CDM model, consisting of the cosmological constant ($\Lambda$) and cold dark matter (CDM), is the standard model that consistently fits the current observational data sets. However, the cosmological constant suffers from fine-tuning and coincidence issues and in order to address these issues, modified matter models like coupled dark energy, unified dark energy, quintessence, and k-essence are suggested. The variable time dependency of the EoS parameter ($\omega$) provides more physically viable and realistic dark energy models, which are probed by the CPL (Chevalliear-Polarski-Linder) parametrization \cite{Chevallier01, Linder03}.

Late-time accelerated expansion can also be studied by modified gravitational theory without using the dark energy model. However, this approach is impossible in GR \cite{Lue06, Nojiri07a}, and one can extend the geometrical part of the Einstein-Hilbert action to address the cosmic expansion issue. In the teleparallel formulation of gravity, higher curvature corrections can be introduced such as the Gauss-Bonnet combination $\mathcal{G}$, so the action would involve higher-torsion modifications \cite{Boulware85, Wheeler86, Antoniadis94, Nojiri05a}. The torsion invariant $T_\mathcal{G}$ has been extracted without imposing the Weitzenb$\ddot{o}$ck connection, equivalent to the Gauss-Bonnet term $\mathcal{G}$ \cite{Kofinas14}. This has led to another interesting class  of modified theories of gravity, known as $F(T, T_\mathcal{G})$ gravity \cite{Kofinas14, Kofinas14a}. Another modified gravity formulated with the torsion scalar $T$ is coupled with the trace of energy-momentum tensor $\mathcal{T}$. In the cosmological applications, the unified description of the inflationary phase, matter-dominated expansion, and late time acceleration can be realized \cite{Harko14a}. Also, the extension of $f(T)$ gravity can be obtained by including the non-minimal torsion-matter coupling in the action of $f(T)$ gravity. This has been successful in getting the dark energy sector of the Universe \cite{Harko14b}. This study explores a gravitational action composed of the torsion scalar and the Gauss-Bonnet component, leading to the $F(T, T_\mathcal{G})$ theories. These have been extensively studied in various contexts (see, for example, \cite{Kofinas14b, Chattopadhyay14, Capozziello16}), yielding exciting results on multiple scales. We focus on the cosmological dynamics of a subclass of $F(T, T_\mathcal{G})$ models chosen based on symmetry considerations. Our goal is to use late-time cosmic observations to test the viability of such a scenario and determine whether it could be a viable alternative to the standard cosmological paradigm.

Motivated by the successful cosmological results of the extension of $f(T)$ gravity, we shall study the cosmological scenario in $F(T, T_\mathcal{G})$ gravity in this paper. In particular, we shall focus on the behavior of the Universe at the late time of its evolution. The article is organized as follows: In Sec. \ref{SEC II}, we have set up the field equations of $F(T, T_\mathcal{G})$ gravity. In Sec. \ref{SEC III}, we have performed observational constraints using Hubble and Pantheon data. The $F(T, T_\mathcal{G})$ model is suggested to obtain the solutions to the field equations, including the behavior of cosmological parameters such as deceleration parameter and EoS parameter and also discussed energy conditions in Sec. \ref{SEC IV}. In Sec. \ref{SEC V}, we present the $Om(z)$ diagnostic and the age of the Universe. Finally, Sec. \ref{SEC VI}, we have presented the results and conclusions. 

\section{$F(T,T_\mathcal{G})$  gravity  Field Equations and Dynamical Parameters} \label{SEC II}
In $F(T, T_\mathcal{G})$ gravity, the total modified gravitational action has the following form \cite{Kofinas14}:
\begin{equation}\label{eq.1}
S=\frac{1}{2\kappa^2} \int d^{4}x\,\, e\,\, F(T, T_\mathcal{G}),
\end{equation}
which is based on the torsion scalar $T$ and the Gauss-Bonnet invariant term, $T_\mathcal{G}$. We consider $\kappa^2=8 \pi G$ with $G$ denoting the Newtonian gravitational constant. In curvature-based gravity, it is possible to express the Gauss-Bonnet invariant as,
\begin{equation*}
\mathcal{G} \equiv R^2-4R^{a b} R_{a b}+R^{a b c d}R_{a b c d}    
\end{equation*}
where $R$, $R^{a b}$, $R^{a b c d}$ denotes the Ricci scalar, Ricci tensor, and Riemann tensor respectively.
Whereas in the torsion-based gravity, $F(T, T_{\mathcal{G}})$ gravity, the invariant term $T_\mathcal{G}$ can be  defined as,
\begin{eqnarray}\label{eq.2}
T_\mathcal{G}=\Big(\mathcal{K}^{a_1}_{~~e a} \mathcal{K}^{e a_2}_{~~b} \mathcal{K}^{a_3}_{~~f c} \mathcal{K}^{f a_4}_{~~d}-2 \mathcal{K}^{a_1 a_2	}_{~~~~a} \mathcal{K}^{a_3}_{~~e b} \mathcal{K}^{e}_{~~f c} \mathcal{K}^{f a_4}_{~~d}+2\mathcal{K}^{a_1 a_2	}_{~~~~a} \nonumber\\\times \mathcal{K}^{a_3}_{~~e b} \mathcal{K}^{e a_4}_{~~f} \mathcal{K}^{f}_{~c d}   +2\mathcal{K}^{a_1 a_2	}_{~~~~a} \mathcal{K}^{a_3}_{~~e b} \mathcal{K}^{e a_4}_{~~c,d} \Big)\delta^{a b c d}_{a_1 a_2 a_3 a_4}.~~~~~~~
\end{eqnarray}
The following gravitational field equations obtained by modifying the action \eqref{eq.1} about vierbein:
\begin{eqnarray}\label{eq.3}
&& 2(H^{[ac]b}+H^{[ba]c}-H^{[cb]a})_{,c} +2(H^{[ac]b}+H^{[ba]c}-H^{[cb]a}) \mathcal{C}^{d}_{\,\,\,\,dc}\nonumber\\ &&+(2H^{[ac]d}+H^{dca}) \mathcal{C}^{b}_{\,\,\,\,cd}+4 H^{[db]c} \mathcal{C}_{(dc)}^{\,\,\,\,\,\,\,\,\,a}+T^{a}_{\,\,\,\,cd} H^{cdb}-h^{ab}\nonumber\\ &&\hspace{0.3cm}+(F-TF_T-T_\mathcal{G} F_{T_\mathcal{G}}) \eta^{ab}=0.
\end{eqnarray}
with 
\begin{eqnarray}
H^{abc}&=&F_T (\eta^{ac} \mathcal{K}^{bd}_{\,\,\,\,\,\,d}-\mathcal{K}^{bca}+F_{T_\mathcal{G}}\Big[\epsilon^{cprt}(2\epsilon^{a}_{\,\,\,dkf} \mathcal{K}^{bk}_{\,\,\,\,\,\,p}\mathcal{K}^{d}_{\,\,\,\,qr} \nonumber \\&&\hspace{-0.3cm}+\epsilon_{qdkf} \mathcal{K}^{ak}_{\,\,\,\,\,\,p} \mathcal{K}^{bd}_{\,\,\,\,\,\,r}+\epsilon^{ab}_{\,\,\,\,\,\,kf} \mathcal{K}^{k}_{\,\,\,\,dp} \mathcal{K}^{d}_{\,\,\,\, qr}) \mathcal{K}^{qf}_{\,\,\,\,\,\,t}
  + \epsilon^{cprt} \nonumber \\&& \hspace{-0.3cm}\times \epsilon^{ab}_{\,\,\,\,kd} \mathcal{K}^{fd}_{\,\,\,\,\,\,p} (\mathcal{K}^{k}_{\,\,\,\,fr,t}-\frac{1}{2} \mathcal{K}^{k}_{\,\,\,\,fq} \mathcal{C}^{q}_{\,\,\,tr})+\epsilon^{cprt}\epsilon^{ak}_{\,\,\,\,\,\,df} \mathcal{K}^{df}_{\,\,\,\,\,\,p}\nonumber \\&& \hspace{-0.3cm}\times (\mathcal{K}^{b}_{\,\,\, kr,t}-\frac{1}{2} \mathcal{K}^{b}_{\,\,\,\, kq} \mathcal{C}^{q}_{\,\,\,\,tr})\Big]+\epsilon^{cprt} \epsilon^{a}_{\,\,\,\,kdf}\big[(F_{T_\mathcal{G}} \mathcal{K}^{bk}_{\,\,\,\,\,\,p} \mathcal{K}^{df}_{\,\,\,\,\,\,\,r})_{,t}\nonumber \\&&\hspace{-0.3cm}+F_{T_\mathcal{G}} \mathcal{C}^{q}_{\,\,\,\,pt} \mathcal{K}^{bk}_{\,\,\,\,\,\,[q} \mathcal{K}^{df}_{\,\,\,\,\,\,r]}\big]
\end{eqnarray}
and
\begin{equation*}
    h^{ab}=F_T \epsilon^{a}_{\,\,\,\,kcd} \epsilon^{bpqd} \mathcal{K}^{k}_{\,\,\,\,fp} \mathcal{K}^{fc}_{\,\,\,\,\,\,q}
\end{equation*}
where $F_T$ and $F_{T_\mathcal{G}}$ respectively denote the partial derivative with respect to the torsion scalar $T$ and Gauss-Bonnet invariant $T_\mathcal{G}$.
To derive the field equations of $F(T,T_{\mathcal{G}})$, we consider an isotropic and homogeneous FLRW space-time as, 
\begin{equation} \label{eq.4}
ds^{2}=-dt^{2}+a^{2}(t)(dx^{2}+dy^{2}+dz^{2}),
\end{equation}
where $a(t)$ is the scale factor that measures the expansion rate of the Universe along the spatial directions, which appears to be uniform. For such space-time, the diagonal vierbein is,
\begin{equation}\label{eq.5}
e^{a}_{\,\,\, b}=\text{diag}(1, a(t), a(t), a(t))
\end{equation} 
and its determinant is $e=a^3(t)$, where its dual is represented as 
\begin{equation} \label{eq.6}
e^{\,\, \, b}_{a}=\text{diag}(1, a^{-1}(t), a^{-1}(t), a^{-1}(t))
\end{equation}
Now, the torsion scalar and Gauss-Bonnet invariant term can be expressed respectively in Hubble term as, $T=6H^{2}$ and $ T_{\mathcal{G}}=24H^{2}(\dot{H}+H^{2})$, where $\left(H=\frac{\dot{a}(t)}{a(t)}\right)$, the derivative with respect to cosmic time $t$ is shown over the dot. In addition, we consider a matter action $\mathcal{S}_m$, which is equivalent to an energy-momentum tensor $\mathcal{T}^{ab}$, with a particular emphasis on the case of a perfect fluid with energy density $\rho$ and pressure $p$. 

Varying the total action $\mathcal{S} + \mathcal{S}_m$, the following equations are produced in FLRW geometry \cite{Kofinas14,Kofinas14a}:

\begin{eqnarray} 
&& F-12 H^2 F_{T} - T_{\mathcal{G}} F_{T_{\mathcal{G}}}+24 H^3 \dot{F}_{T_{\mathcal{G}}}=2 \kappa^2 \rho,  \label{eq.7} \\
&& F-4(\dot{H}+3 H^2)F_{T}-4 H \dot{F}_{T}-T_{\mathcal{G}} F_{T_{\mathcal{G}}}+\frac{2}{3H} T_{\mathcal{G}} \dot{F}_{T_{\mathcal{G}}}\nonumber\\&&\hspace{4.5cm}+8 H^2 \ddot{F}_{T_{\mathcal{G}}}=-2 \kappa^2 p.   \label{eq.8}
\end{eqnarray} 
with
\begin{eqnarray} \label{eq.9}
F_T \equiv \frac{\partial F(T, T_\mathcal{G})}{\partial T},\,\,\,\,\,\,\,\,\,\,\,\,\, F_{T_\mathcal{G}} \equiv \frac{\partial F(T, T_\mathcal{G})}{\partial T_\mathcal{G}}
\end{eqnarray}
For brevity, we represent $F\equiv F(T, T_{\mathcal{G}})$. To frame the cosmological model, we calculate the pressure and energy density for a physically acceptable form of $F(T, T_\mathcal{G})$. Now, the derivative of $F$ can be obtained as,
\begin{eqnarray}
&&\hspace{-0cm}\dot{F}_T=F_{TT}\dot{T}+F_{TT_{\mathcal{G}}}\dot{T}_{\mathcal{G}} \nonumber\\
&&\hspace{-0cm}\dot{F}_{T_{\mathcal{G}}}=F_{T T_\mathcal{G}}\dot{T}+F_{T_\mathcal{G} T_\mathcal{G}}\dot{T}_\mathcal{G} \nonumber\\ &&\hspace{-0cm}\ddot{F}_{T_\mathcal{G}}=F_{TTT_{\mathcal{G}}}\dot{T}^2+2F_{{T T_\mathcal{G}} T_\mathcal{G}}\dot{T}\dot{T}_\mathcal{G}+F_{T_\mathcal{G} T_\mathcal{G} T_\mathcal{G}}\dot{T}_{\mathcal{G}}^2+F_{T T_{\mathcal{G}}} \ddot{T}\nonumber\\&&\hspace{6cm}+F_{T_\mathcal{G} T_\mathcal{G}}\ddot{T}_{\mathcal{G}} ,\nonumber   
\end{eqnarray}
where $F_{TT}$, $F_{TT_\mathcal{G}}$,...are the mathematical expression used to indicate several partial differentiations of $F(T, T_\mathcal{G})$ over $T$, $T_\mathcal{G}$.

\section{Observational Constraints} \label{SEC III}
We use the MCMC sampling method, and the Python emcee \cite{Foreman13} package to explore the parameter space. To note, the normalizing constant will not be computed to estimate the parameters. However, the prior and likelihood estimate can be used to calculate the posterior parameter distributions. In this analysis, we have used the recently released Pantheon dataset, which contains 1048 Supernova Type Ia experiment findings from surveys such as the Low-z, SDSS, SNLS, Pan-STARRS1(PS1) Medium Deep Survey, and HST \cite{Scolnic18}, in the redshift range $z\in(0.01,2.26)$. With an emphasis on the evidence relevant to the expansion history of the Universe, such as the distance-redshift connection, two separate current observational datasets are employed to limit the model under consideration.  More importantly, new studies investigating the roles of $H(z)$ and SNeIa data in cosmological constraints have found that both can restrict cosmic parameters. The parameters for this model are $\alpha$, $\beta$, $\zeta,$ and $\Omega_{m0}$. To determine the expansion rate $(1+z) H(z)=-\frac{dz}{dt}$, we rewrite $T$ and $T_\mathcal{G}$ in redshift parameter as,
\begin{small}
\begin{eqnarray}
 \label{eq.10}
    T=6 H_0^2 E(z),~~ T_\mathcal{G}=24 H_0^2 E(z) \left(- \frac{H_0^2 (1+z) E'(z)}{2}+H_0^2 E(z) \right),~~~~ 
\end{eqnarray}
\end{small}
where $H^2 (z) = H_0^2 E(z)$ and $H_0 = 67 \pm 4~km\,s^{-1} Mpc^{-1}$\cite{Yu18} be the late-time Hubble parameter and the prime denotes the derivative to the redshift parameter. In addition, we have considered $H_0=70.7$ $kms^{-1} Mpc^{-1}$ for our analysis. We use the following functional form for $E(z)$ [Sahni et al.\cite{Sahni03}],
\begin{equation}\label{eq.11}
    E(z)=\Omega_{m0} (1+z)^3 + \zeta (1+z)^2 + \beta (1+z)+ \alpha,
\end{equation}
where the constants $\Omega_{m0}$, $\zeta$, $\beta$, and $\alpha$ are determined by fitting the experimental data and their measurements. Additionally, the restriction $E(z=0)=1$ constrains the relationship between these coefficients, as $\alpha + \beta + \zeta =1-\Omega_{m0}$.
\subsection{Hubble Data}
To recreate the cosmological models, we employed the parametrization approach. Some interesting works have utilized the parametrization approach to investigate cosmological models \cite{Saini00, Capozziello05}. The real benefit of using this approach is that observational data may be used to evaluate cosmological theories. It is well recognized that the SNe Ia has a solid ability to restrict cosmological theories. However, it is challenging to depict the precise history of $H$ due to the integration in its formula $H(z)$. As a result, the $H(z)$ data can indicate the fine structure of the expansion history of the Universe. Sharov et al. \cite{Sharov18} present the entire list of data sets. We estimate the model parameters using the $\chi^2$ test using MCMC simulation. The $\chi_{\text{Hubble}}^2$ value for the  observational Hubble parameter data can be expressed as 
\begin{equation} \label{eq.12}
    \chi_{\text{Hubble}}^{2}=\sum_{i=1}^{55}\frac{\left[H_{th}(z_i)-H_{\text{obs}}(z_i)\right]^2}{\sigma_{i}^{'2}},
\end{equation}
where $\sigma_{i}^{'}$ denotes the standard error in experimental values of the Hubble function. The notations $H_{th}(z_i)$ and $H_{\text{obs}}(z_i)$ represent the theoretical and observable values of the Hubble parameter $H$, respectively.

\begin{figure*} [!htb]
\centering
\includegraphics[width=8.9cm,height=6cm]{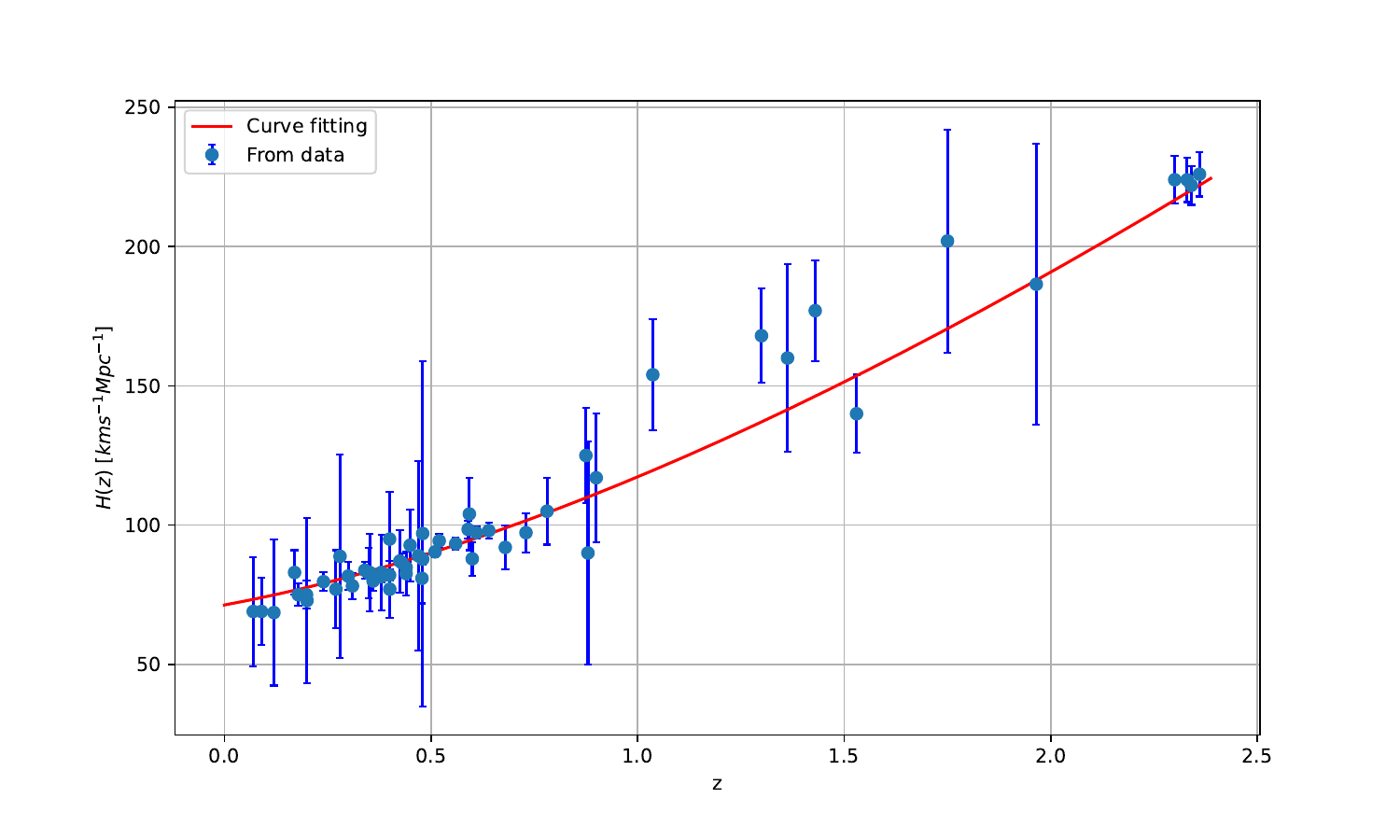} \includegraphics[width=8.9cm,height=6cm]{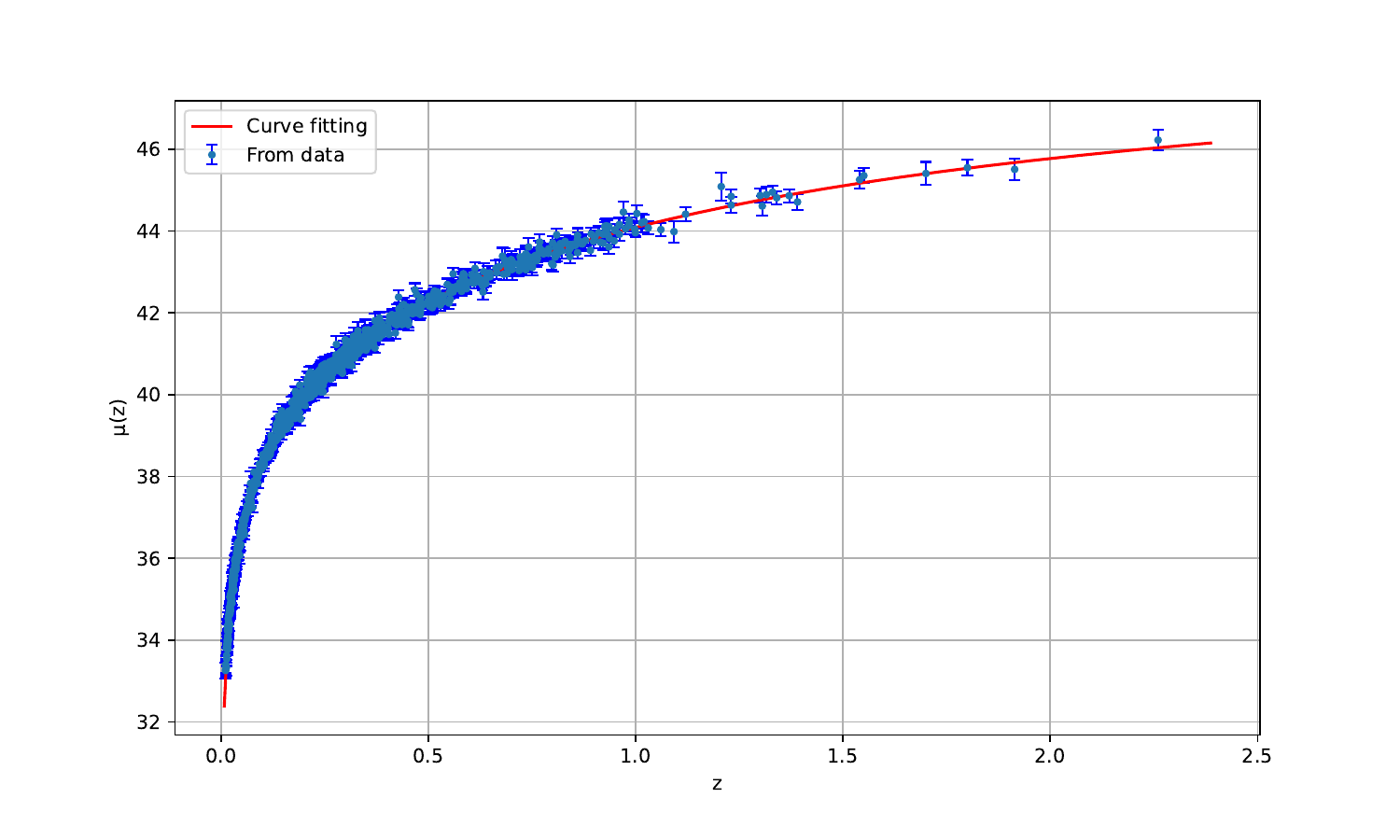}
\caption{The provided model (solid red line) has a better fit to the $H(z)$ datasets for $\alpha=0.721$, $\beta=0.030$, $\zeta=0.043$ and $\Omega_{m0}=0.226$, which is shown in the left panel plot along with the 55 points of the $H(z)$ datasets (blue dots) and accompanying error bars (see Table \ref{table: Table II}). In the right panel, The red line is the plot of our model's distance modulus $\mu(z)$ vs $z$, which exhibits a better fit to the 1048 points of the Pantheon datasets along with its error bars for $\alpha=0.716$, $\beta=0.024$, $\zeta=0.024$ and $\Omega_{m0}=0.2599$.}
\label{FIG1}
\end{figure*}

\subsection{Pantheon Data}
In a sample of 1048 SNe Ia from the Pantheon study, the $\chi^2_{\text{Pantheon}}$ function is provided by \cite{Scolnic18},
\begin{equation} \label{eq.13}
    \chi_{\text{Pantheon}}^{2}=\sum_{i=1}^{1048}\frac{\left[\mu_{th}(\mu_0,z_i)-\mu_{\text{obs}}(z_i)\right]^2}{\sigma_{i}^{'2}}
\end{equation}

Furthermore, the standard error in the practical value of $H$ is indicated by the symbol $\sigma_{i}^{'}$. The theoretical distance modulus $\mu_{th}$ is defined by $\mu_{th}^i=\mu(D_{L})=m-M=5 log_{10}D_L(z)+\mu_0$, where the apparent and absolute magnitudes are denoted by $m$ and $M$, respectively, and the nuisance parameter $\mu_0$ is specified as $\mu_0=5log(\frac{H_0^{-1}}{Mpc})+25$. The formula for luminosity distance $D_L$ is given by $D_L(z)=(1+z) H_0 \int \frac{1}{H(z^*)} dz^*$. The series of $H(z)$ is constrained to the tenth term and approximately integrates the constrained series to get the luminosity distance.

\begin{figure*} [!htb]
\centering
\includegraphics[width=120mm]{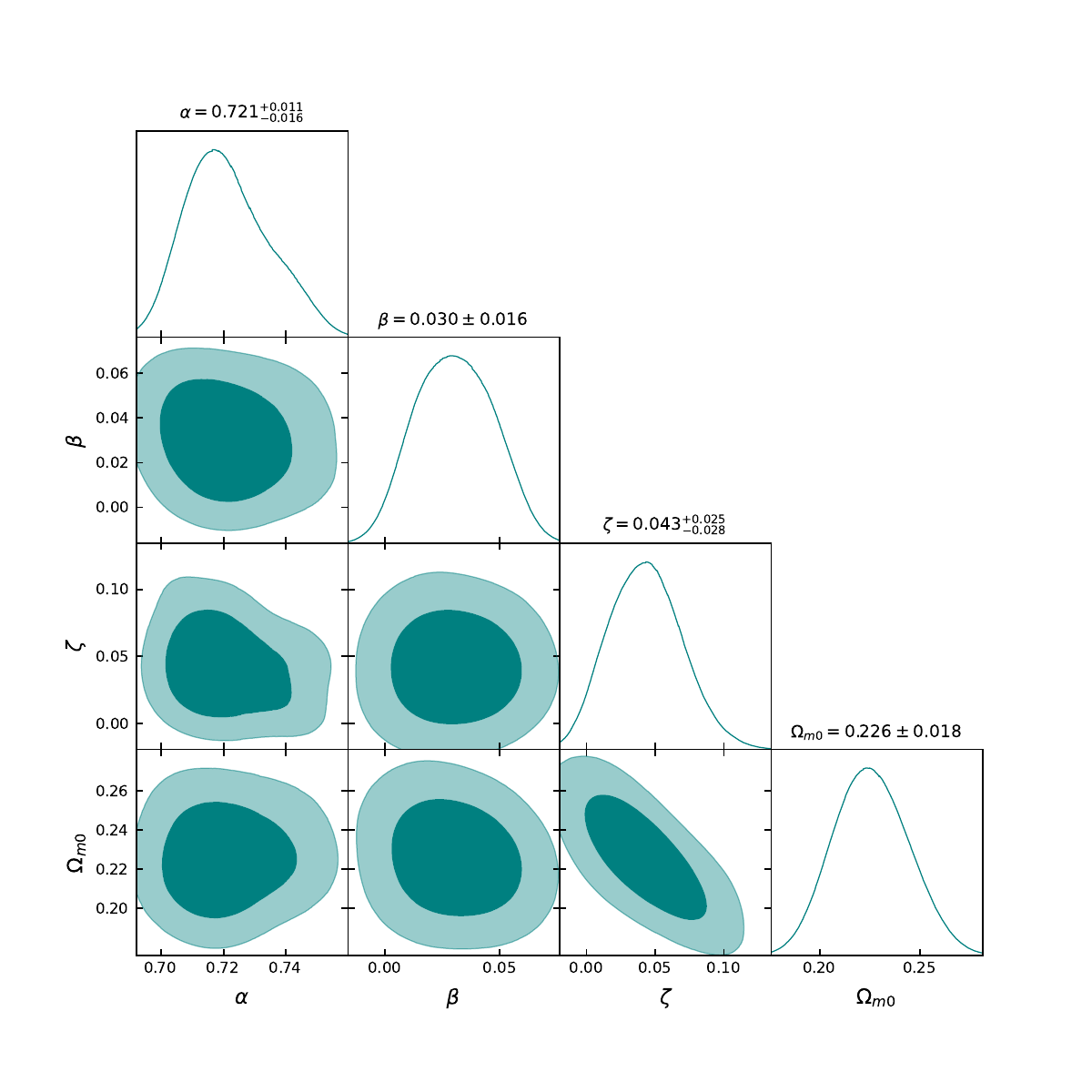}
\caption{The contour plots with $1-\sigma$ and $2-\sigma$ errors for the parameters $\alpha$, $\beta$, $\zeta$, and $\Omega_{m0}$. Additionally, it contains the parameter values that better match the 55-point Hubble dataset defined in Table \ref{table: Table II}.}
\label{FIG2}
\end{figure*}
\begin{figure*} [!htb]
\centering
\includegraphics[width=120mm]{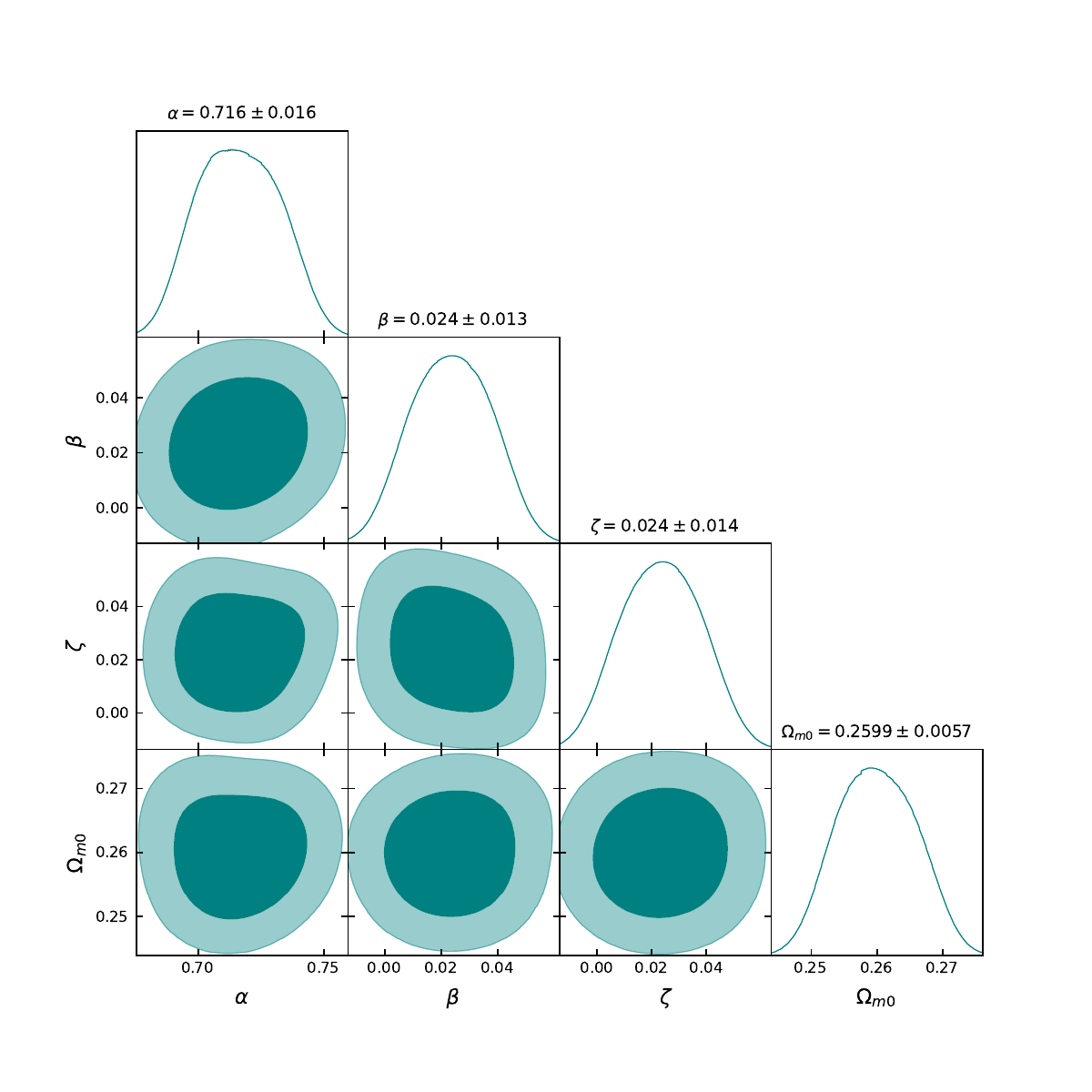}
\caption{The contour plots with $1-\sigma$ and $2-\sigma$ errors for the parameters $\alpha$, $\beta$, $\zeta$, and $\Omega_{m0}$. It also contains parameter values that better match the 1048-point Pantheon sample.}
\label{FIG3}
\end{figure*}

Fig. \ref{FIG1} (left panel) shows the behaviour of Hubble parameter from $H(z)$ data set and Fig. \ref{FIG1} (right panel) shows the behaviour of $\mu(z)$ from 1048 points of the Pantheon data set. In Fig. \ref{FIG2} and Fig. \ref{FIG3}, we see the marginalized distribution for the parameters $\alpha$, $\beta$, $\zeta$ and $\Omega_{m0}$ which has been displayed with the triangle plots. The contour shows where the 1-$\sigma$ and 2-$\sigma$ confidence intervals are located.

\begin{table} [!htb]
\caption{Constraining Parameters} 
\centering 
\begin{tabular}{|c|c|c|} 
\hline\hline 
Coefficients & Hubble Dataset  & Pantheon Dataset\\ [0.5ex] 
\hline
$\alpha$ & $0.721^{+0.011}_{-0.016}$ & 0.716 $\pm$ 0.016 \\
\hline
$\beta$ & 0.030 $\pm$ 0.016 & 0.024 $\pm$ 0.013 \\
\hline
$\zeta$ & $0.043_{-0.028}^{+0.025}$ & 0.024 $\pm$ 0.014 \\
\hline
$\Omega_{m0}$ & 0.226 $\pm$ 0.018  & 0.2599$\pm$0.0057 \\[0.5ex] 
\hline 
\end{tabular}
\label{table:nonlin} 
\label{TABLE I}
\end{table}
Fig. \ref{FIG2} and Fig. \ref{FIG3}  exhibit the 1-$\sigma$ and 2-$\sigma$ confidence regions that have been illustrated in our constraint findings. These are retrieved by the respective contour analyses of $\chi^2$ in the parameter space. Table \ref{TABLE I} also summarizes the best-fit value of the parameters and their associated uncertainty.\\

\section{The Functional $F(T,T_\mathcal{G})$} \label{SEC IV}
The above analysis requires the specification of the $F(T, T_\mathcal{G})$ form. The corrections of $T$-powers are included first in conventional $f(T)$ gravity. However, in the present context, $T_\mathcal{G}$ is in the same order as $T^2$ because it includes quartic torsion components. Because $T$ and $\sqrt{T^2 +\lambda_2 T_\mathcal{G}}$ have the same order, both should be employed in a modified theory. As a result, the most fundamental non-trivial model, which is distinct from GR and does not introduce a new mass scale into the problem, is $F(T, T_\mathcal{G})=-T + \lambda_1 \sqrt{T^2+\lambda_2 T_\mathcal{G}}$. The couplings $\lambda_1, \lambda_2$ are dimensionless, and the model is predicted to be essential in late times. Although straightforward, this model can produce remarkable cosmic behavior demonstrating the advantages, possibilities, and novel aspects of $F(T, T_\mathcal{G})$ cosmology. We note here that this scenario simplifies to TEGR, or GR, with simply a rescaled Newton's constant, whose dynamical analysis has been carried out in detail in the literature \cite{Copeland98, Ferreira97, Chen09} when $\lambda_2 = 0$. Therefore, in the following sections, we restricted our study to the condition $\lambda_2 \neq 0$.

Now plugging the Hubble parameter $H(z)$ and $F(T, T_\mathcal{G})$ model into the field Eqns. \eqref{eq.7} and \eqref{eq.8} the following set of field equations are obtained, 
\begin{widetext}
\begin{eqnarray}
\rho&=& -\frac{T}{2}+\frac{\lambda_1  \left(2 T^2+\lambda_2 T_\mathcal{G} \right)}{4 \sqrt{T^2+\lambda_2 T_\mathcal{G}}}+H^2 \left(6-\frac{6 \lambda_1  T}{\sqrt{T^2+\lambda_2 T_\mathcal{G}}}\right)-\frac{3 \lambda_1  \lambda_2  H^3 \left(2 T \dot{T}+\lambda_2  \dot{T}_\mathcal{G}\right)}{\left(T^2+\lambda_2 T_\mathcal{G}\right)^{3/2}} \label{eq.15}\\
p&=& \frac{T}{2}-\frac{\lambda_1  \sqrt{T^2+\lambda_2 T_\mathcal{G}}}{2}+\frac{\lambda_1  \lambda_2 T_\mathcal{G}}{4 \sqrt{T^2+\lambda_2 T_\mathcal{G}}}+2 \left(\dot{H}+3 H^2\right) \left(\frac{\lambda_1  T}{\sqrt{T^2+\lambda_2 T_\mathcal{G}}}-1\right)-\frac{ \lambda_1  \lambda_2  H \left(T \dot{T}_\mathcal{G}-2 T_\mathcal{G} \dot{T}\right)}{\left(T^2+\lambda_2  T_\mathcal{G}\right)^{3/2}}\nonumber\\& &+\frac{\lambda_1 \lambda_2 T_\mathcal{G} \left(2 T \dot{T}+\lambda_2 \dot{T}_\mathcal{G}\right)}{12 H \left(T^2+\lambda_2 T_\mathcal{G}\right)^{3/2}}-\frac{\lambda_1  \lambda_2  H^2 \left(3 \left(2 T \dot{T}+\lambda_2  \dot{T}_\mathcal{G}\right)^2-2 \left(T^2+\lambda_2 T_\mathcal{G}\right) \left(2 T \ddot{T}+2 \dot{T}^2+\lambda_2  \ddot{T}_\mathcal{G}\right)\right)}{2 \left(T^2+\lambda_2 T_\mathcal{G}\right)^{5/2}}  \label{eq.16}
\end{eqnarray}
\end{widetext}
where, $T=6 H^2 ,\hspace{0.2cm} T_{\mathcal{G}}=24H^{2}(\dot{H}+H^{2}),\hspace{0.2cm} \dot{T}=12 H \dot{H}, \hspace{0.2cm} \dot{T}_\mathcal{G}= 24 H \left(H \ddot{H}+2 \dot{H} (\dot{H}+2 H^2)\right), \hspace{0.2cm} \ddot{T}=12 H \ddot{H}+12 \dot{H}^2,\\ \ddot{T}_\mathcal{G}= 24 \left(4 H^3 \ddot{H}+2 \dot{H}^3+H^2 (\dot{\ddot{H}}+12 \dot{H}^2)+6 H \dot{H} \ddot{H}\right)$ and by using $(1+z) H(z)=-\frac{dz}{dt}$, we obtained $\dot{H}, \ddot{H}$ and $\dot{\ddot{H}}$ over redshift as mentioned in {\bf Appendix}.

\subsection{Deceleration parameter and EoS parameter}
\begin{figure*} [!htp]
\centering
\minipage{1\textwidth}
\includegraphics[width=\textwidth]{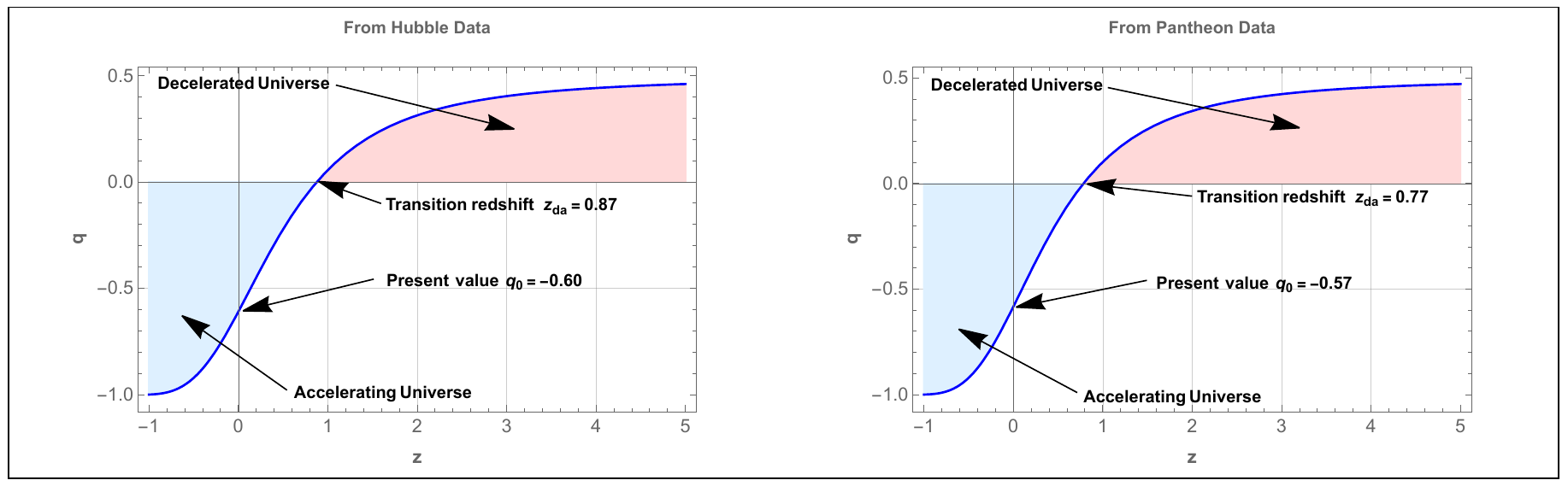}
\endminipage
\caption{Graphical behavior of the deceleration parameter versus redshift with the constraint values of the coefficients obtained from Fig. \ref{FIG2} and \ref{FIG3} (The parameter scheme: Mean of parameter values).}
\label{FIG4}
\end{figure*}
\begin{figure*} [!htp]
\centering
\minipage{1\textwidth}
\includegraphics[width=\textwidth]{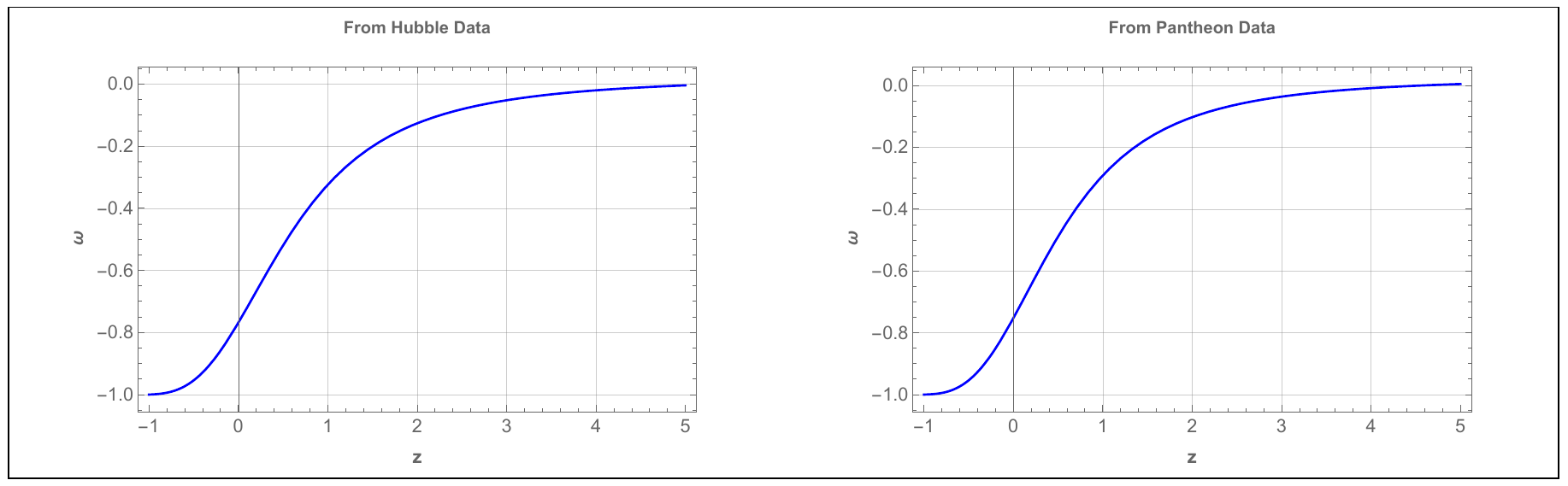}
\endminipage
\caption{Graphical behavior of the EoS parameter versus redshift with the constraint values of the coefficients obtained from Fig. \ref{FIG2} and \ref{FIG3} (The parameter scheme: Mean of parameter values).}
\label{FIG5}
\end{figure*}
The deceleration parameter $q=-1-\frac{\dot{H}}{H^2}$, is a function of the Hubble parameter that describes the rate of acceleration of the Universe. For positive $q$, the Universe is in a decelerated phase; for negative $q$, the accelerated phase can be realized. The model parameters $\alpha, \beta, \zeta,$ and $\Omega_{m0}$ are used to calculate the deceleration parameter $q$. The graph explains the expansion from the past to the present by depicting how $q$ behaves for redshift $z$. In Fig. \ref{FIG2} and Fig. \ref{FIG3}, the restricted values of model parameters from the examined Hubble and Pantheon data sets cause $q$ to transit from positive in the past, indicating early deceleration, to negative in the present, indicating current acceleration. The deceleration parameter $q_0=-0.60$, $q_0=-0.57$ for Hubble and Pantheon data respectively, at the current cosmic epoch, is relatively consistent with the range $q_0=-0.528^{+0.092}_{- 0.088}$ as determined by recent observation \cite{Gruber14}.

The Universe makes a smooth transition from a decelerated phase of expansion to an accelerated phase in our derived model, with a deceleration-acceleration redshift of $z_{da} = 0.87$, $z_{da}=0.77$ for Hubble and Pantheon data respectively shown in Fig. \ref{FIG4}. The recovered value of the deceleration-acceleration redshift $z_{da}=0.82 \pm 0.08$ is consistent with certain current constraints, based on 11 $H(z)$ observations made by Busca et al. \cite{Busca13} between redshifts $0.2 \leq z \leq 2.3$, $z_{da}=0.74\pm 0.05$ of Farooq et al. \cite{Farooq13}, $z_{da}=0.69^{+23}_{-12}$ of Lu et al. \cite{Lu11}, $z_{da}=0.7679^{+0.1831}_{-0.1829}$ of Capozziello et al. \cite{Capozziello14}, and $z_{da}=0.60^{+0.21}_{-0.12}$ of Yang et al. \cite{Yang20}.\\

The kinematic variables are significant in the analysis of cosmological models. The deceleration parameter, for instance, defines the behavior of the Universe, including whether it is always decelerating, constantly accelerating, has a single or several transition phases, etc. The EoS parameter similarly defines the physical significance of energy sources in the evolution of the Universe. The EoS parameter $(\omega)$ is,
\begin{equation}\label{eq.18}
\omega= \frac{p}{\rho}.
\end{equation}
In the dust phase, the EoS parameter, $\omega= 0$, whereas in the radiation-dominated phase, $\omega = \frac{1}{3}$. The vacuum energy or the $\Lambda$CDM model is represented by $\omega=-1$. In addition, for the accelerating phase of the Universe, e.g. in the quintessence phase $(-1 < \omega < 0)$ and in  phantom regime $(\omega < -1)$.

We may visualize the variations in EoS of dark energy Eq. \eqref{eq.18} in terms of the redshift variable by calculating the associated energy density and pressure of dark energy, as shown in Fig. \ref{FIG5}. This diagram represents the quintessence-like behavior and its approach to $-1$ at late times, so that the current value of EoS $(z = 0)$ equals $-0.77$, $-0.755$ for Hubble and Pantheon data respectively for the values of model parameter $\lambda_1=0.3, \lambda_2=0.36$. As a result, we conclude that the Universe is expanding faster, which is compatible with the cosmological data provided by Amanullah et al. \cite{Amanullah10}.

The Pantheon study constrained the parameter space, a newly proposed observational dataset. The $2-\sigma$ limitations for the parameters in our study are $\alpha =0.716 \pm 0.016, \beta =0.024 \pm 0.013$, $\zeta=0.024 \pm 0.014$ and $\Omega_{m0} = 0.2599 \pm 0.0057$. The 1048 Pantheon samples and our model taking into account $H_0 = 70.7$ kms$^{-1}$ Mpc$^{-1}$, show a good fit to the observational findings in the error bar plot. Valentino et al. \cite{Valentino16} have performed the combined analysis of the Planck and $R16$ results in an extended parameter space. In place of the usual six cosmological parameters, twelve parameters were simultaneously varying and obtained the phantom-like dark energy component, with $\omega=-1.29_{-0.12}^{+0.15}$ at $68\%$ of C.I. Some other experiments on this parameter suggests the range for EoS parameter as $\omega\approx -1.3$ \cite{Vagnozzi20}. In addition, Efstathiou and Gratton \cite{Efstathiou20} have obtained the range of the curvature density parameter $\Omega_k=0.0004\pm0.0018$, which is in agreement with the Planck 2018 result. Further Vagnozzi et al. \cite{Vagnozzi21} obtained $\Omega_k=0.0054\pm0.0055$, which is consistent with the spatially flat Universe by combining Planck 2018 CMB temperature and polarization data with the latest cosmic chronometer measurements. 

\subsection{Energy Conditions}
In defining energy conditions, the well-known Raychaudhuri equation, which deals with attractive gravity, has proved to be very useful. The Raychaudhuri equation indicates that \cite{Santos07, Kar07},
\begin{equation} \label{eq.19}
\frac{d\theta}{d\tau}=-\frac{1}{2} \theta^2-\sigma_{ab} \sigma^{ab}+w_{ab} w^{ab}-R_{ab} k^a k^b,
\end{equation}
\begin{figure*} [!htb]
\centering
\minipage{1\textwidth}
\includegraphics[width=\textwidth]{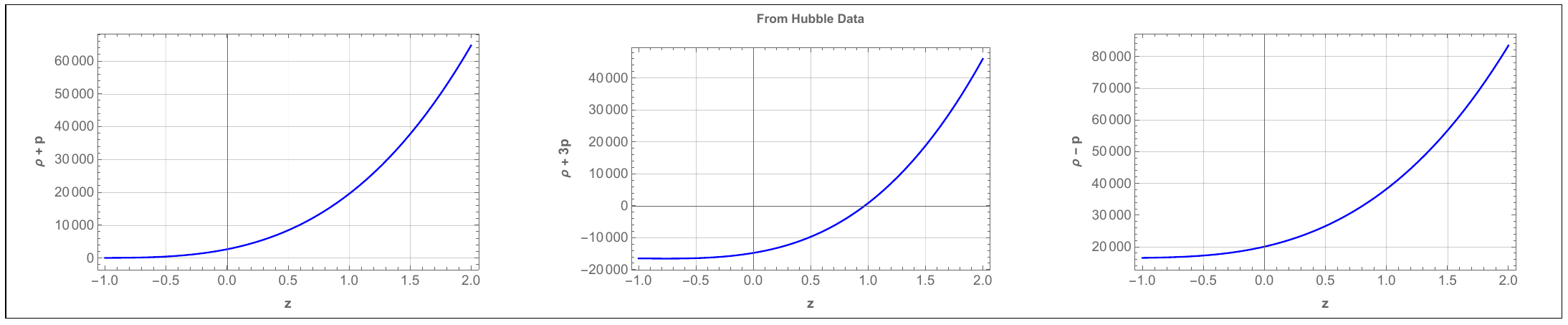}
\endminipage
\caption{Graphical behavior of the energy conditions versus redshift with the constraint values of the coefficients obtained from Fig. \ref{FIG2} and \ref{FIG3} (The parameter scheme: Mean of parameter values).}
\label{FIG6}
\end{figure*}
\begin{figure*} [!htb]
\centering
\minipage{1\textwidth}
\includegraphics[width=\textwidth]{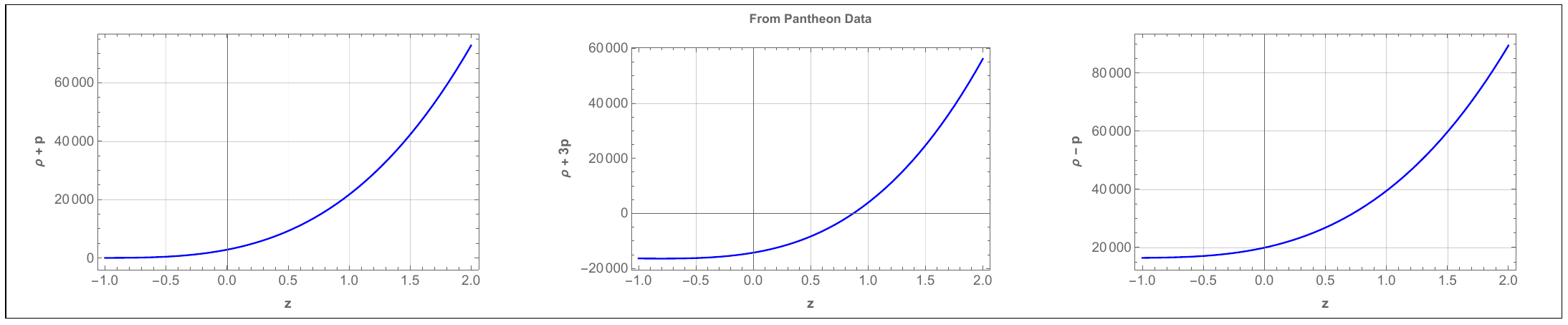}
\endminipage
\caption{Graphical behavior of the energy conditions versus redshift with the constraint values of the coefficients obtained from Fig. \ref{FIG2} and \ref{FIG3} (The parameter scheme: Mean of parameter values).}
\label{FIG7}
\end{figure*}
where the expansion scalar is $\theta$, the shear and vorticity tensors are $\sigma_{ab}$ and $w_{ab}$, respectively. Also, $k^a$ is a null vector field. The Raychaudhuri equation avoids any reference to gravitational field equations, which is essential to establish. Instead, it is viewed as a purely geometric statement. If we consider any orthogonal congruence hypersurface $(w_{ab}=0)$. Then, as a result of $\frac{d\theta}{d\tau}<0$, we can formulate the criteria for attractive gravity as $R_{ab} k^a k^b \geq 0$ because the shear tensor's spatial nature is $\sigma^2 =\sigma^{ab} \sigma_{a b}\geq 0$. The previous condition, known as the null energy condition, can be written in terms of the stress-energy tensor in the context of Einstein's relativistic field equations as $\mathcal{T}_{ab}k^a k^b \geq 0$, where $k^a$ is any null vector. More precisely, the weak energy condition indicates that $\mathcal{T}_{ab}u^a u^b \geq 0$, where $u^a$ denotes the time-like vector and assumes a positive local energy density.

The energy conditions are essentially boundary conditions for maintaining a positive energy density \cite{Hawking73, Poisson04}. Hence, we present here, Null Energy Condition (NEC): $ \rho+p \geq 0$, Weak Energy Condition (WEC): $\rho \geq 0$ and $\rho + p \geq 0$, Strong Energy Condition (SEC): $ \rho+3p \geq 0$ and $ \rho+p \geq 0$, Dominant Energy Condition (DEC): $\rho \geq 0$ and $\rho \pm p \geq 0$. The NEC violation suggests that none of the energy conditions specified are valid. The SEC is now the topic of significant discussion because of the current accelerated expansion of the Universe \cite{Barcelo02}. SEC must be violated in cosmological situations throughout the inflationary expansion and now \cite{Visser97}. The graph of the energy conditions is shown in Fig. \ref{FIG6} and \ref{FIG7}. We check if the NEC and DEC hold, but the SEC violates the model, which directly points to the accelerated expansion of the Universe.

Fig. \ref{FIG6} and \ref{FIG7} illustrate that the WEC is positive from the early time to the late time phase. Since our model exhibits quintessential behavior, we can predict how satisfied DEC and NEC are at the late stages of evolution. At the same time, the SEC started a violation from $z \approx 0.972$, $z \approx 0.879$, and was previously satisfied for both data sets. Simultaneously, the SEC was violated at the late time from $(z \approx 0.9)$ and satisfied at the early time. In particular, a detailed analysis of these energy conditions may be accomplished when the cosmic dynamics are fixed up by a calculated or assumed Hubble rate.

\section{$Om(z)$ Diagnostic and Age of the Universe} \label{SEC V}
In this section, we are interested in how the model responds to the $Om(z)$ diagnostic. For some dark energy theories, the $Om(z)$ parameter is considered another effective diagnostic tool \cite{Sahni08, Sahni14} and which is defined as,
\begin{figure*} [!htb]
\centering
\minipage{1\textwidth}
\includegraphics[width=\textwidth]{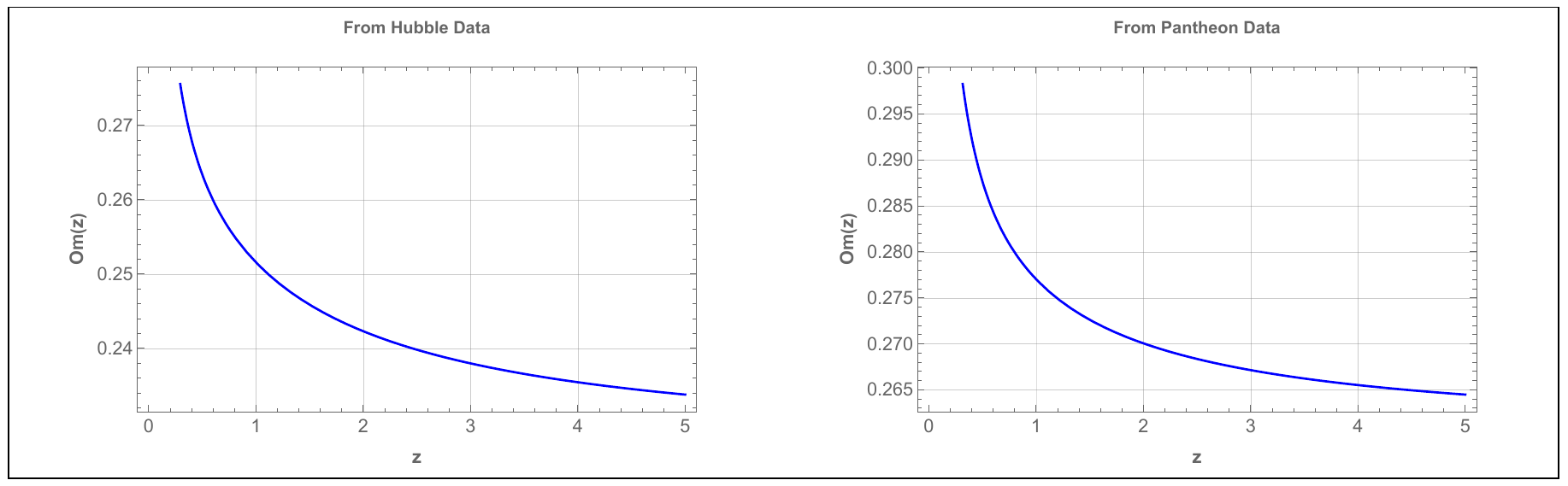}
\endminipage
\caption{Graphical behavior of the $Om(z)$ versus redshift with the constraint values of the coefficients obtained from Fig. \ref{FIG2} and \ref{FIG3} (The parameter scheme: Mean of parameter values).}
\label{FIG8}
\end{figure*}
\begin{figure*} [!htb]
\centering
\minipage{1\textwidth}
\includegraphics[width=\textwidth]{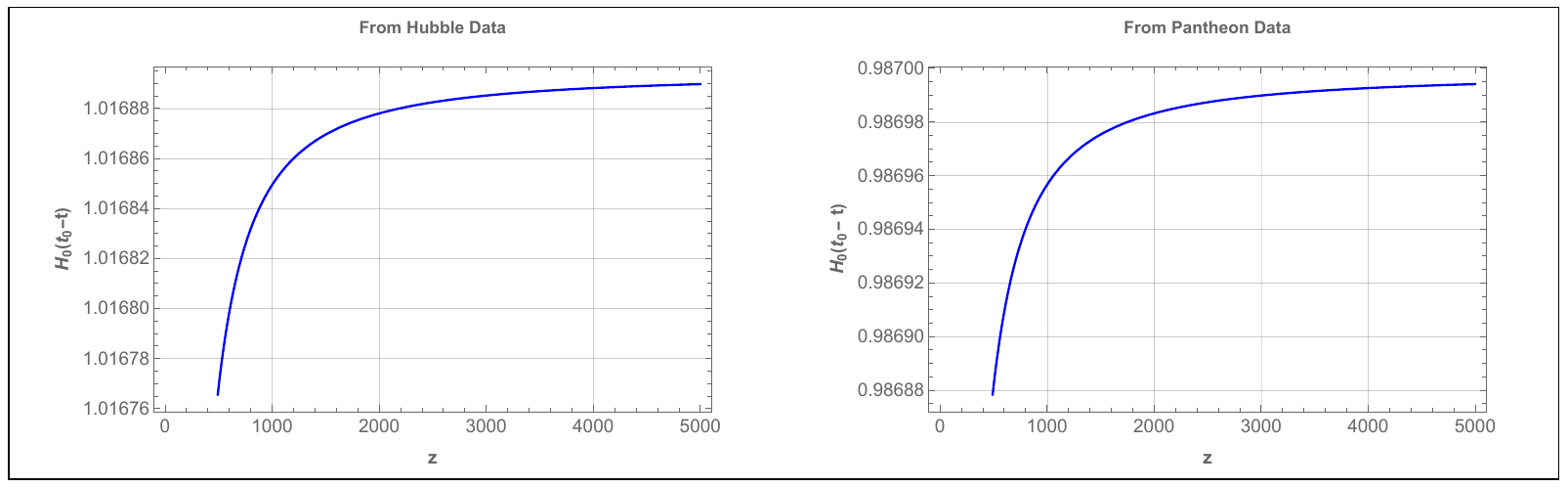}
\endminipage
\caption{Graphical behavior of time versus redshift with the constraint values of the coefficients obtained from Fig. \ref{FIG2} and \ref{FIG3} (The parameter scheme: Mean of parameter values).}
\label{FIG9}
\end{figure*}
\begin{equation} \label{eq.20}
Om(z)=\frac{E(z)-1}{(1+z)^3-1}
\end{equation}
where, $E(z)=\frac{H^2(z)}{H^2_0}$ is dimensionless parameter and $H_0$ is the Hubble rate of the present epoch. The two-point difference diagnostic is
\begin{equation}\label{eq.21}
    Om(z_1, z_2)= Om(z_1) - Om(z_2)
\end{equation}
Alternatively put for quintessence, $Om(z_1,z_2) > 0$, while for phantom  $Om(z_1,z_2)<0$, ($z_1 < z_2$). For the $\Lambda$CDM model, the $Om(z)$ diagnostic provides a null test \cite{Sahni08}, and more data was subsequently gained on its sensitivity with the EoS parameter \cite{Ding15, Zheng16, Qi18}. The dark energy concept will form a cosmological constant if $Om(z)$ is constant for the redshift. The slope of $Om(z)$, which is positive for the emerging $Om(z)$ and denotes phantom phase $(\omega<-1)$ and negative for quintessence region $(\omega > -1)$ also identifies the dark energy models.

The reconstructed $Om(z)$ parameter for the best-fit data is displayed in Fig. \ref{FIG8} as a function of redshift. Over redshift, it has been observed that the $Om(z)$ parameter decreases.

By figuring out the ages of the oldest objects in our galaxy, one can directly estimate the minimum age of the Universe. These are the stars in the Milky Way's galaxy that are metal-poor. The age of the Universe is computed as,
\begin{equation} \label{eq.22}
H_0 (t_0-t)=\int_{0}^{z} \frac{dx}{(1+x) E(x)}\,\,\, , \hspace{0.8cm} E(z)=\frac{H^2(z)}{H^2_0},
\end{equation}
where 
\begin{equation*}
H_0 t_0=\lim_{z\rightarrow \infty} \int_{0}^{z} \frac{dx}{(1+x) E(x)}
\end{equation*}

We may deduce from this straightforward observation that $1/H_0$ should indicate the current age of the Universe, possibly up to a multiplicative factor extremely near to one. The Universe is $13.8$ billion years old, according to observations of the cosmic background radiation \cite{Gribbin15}. Fig. \ref{FIG9} depicts the time behavior with a redshift. It is found that $H_0(t_0-t)$ converges to $1.01689$ and $0.9870$ for Hubble and Pantheon data, respectively, for infinitely large $z$. This translates to $t_0 = 1.01689 H^{-1}_0 \approx 14.01$ Gyrs and $t_0 = 0.987 H^{-1}_0 \approx 13.607$ Gyrs, which is the current age of the Universe and is very near to the age of the Universe calculated from Planck's findings, $t_0 = 13.786\pm 0.020$ Gyrs \cite{Aghanim20}.  It is well known that the age of the Universe at any redshift is inversely proportional to $H_0$. This requires the Universe be older than the oldest objects it contains at any redshift, which will provide an upper limit on $H_0$. Assuming the $\Lambda{CDM}$ model at late times, Vagnozzi et al. \cite{Vagnozzi22}  obtained the $95$ percent confidence level upper limit as, $H_0 < 73.2~~ km/s/Mpc$.
\begin{table*}
\caption{The observational data set that was used in this
paper} 
\centering 
\begin{tabular}{c c c c c | c c c c c | c c c c c} 
\hline\hline 
No. & $z_{i}$ & H(z) & $\sigma_{H}$ & Ref. & No. & $z_{i}$ & H(z) & $\sigma_{H}$ & Ref. &  No. & $z_{i}$ & H(z) & $\sigma_{H}$ & Ref.\\ [0.5ex] 
\hline 
1. & 0.070 & 69.00 & 19.6 & \cite{Zhang14} & 20. & 0.400 & 82.04 & 2.03 &  \cite{Wang17} & 38. & 0.640 & 98.02 & 2.98 &  \cite{Wang17}\\ 
2. & 0.090 & 69.00 & 12.0 & \cite{Jimenez03} & 21. & 0.4004 & 77.00 & 10.20 &  \cite{Moresco16} & 39. & 0.680 & 92.00 & 8.00 & \cite{Moresco12}\\
3. & 0.120 & 68.60 & 26.2 & \cite{Zhang14} & 22. & 0.4247 & 87.10 & 11.20 &  \cite{Moresco16}  & 40. & 0.730 & 97.30 & 7.00 &  \cite{Blake12}\\
4. & 0.170 & 83.00 & 8.00 &  \cite{Simon05}  & 23. & 0.430 & 86.45 & 3.27 &  \cite{Gaztanaga09}  & 41. & 0.781 & 105.0 & 12.00 &  \cite{Moresco12}\\
5. & 0.179 & 75.00 & 4.00 & \cite{Moresco12} & 24. & 0.440 & 82.60 & 7.80 &  \cite{Blake12}  & 42. & 0.875 & 125.0 & 17.00 &  \cite{Moresco12} \\
6. & 0.199 & 75.00 & 5.00 &  \cite{Moresco12} & 25. & 0.440 & 84.81 & 1.83 &  \cite{Wang17}  & 43. & 0.880 & 90.00 & 40.00 &  \cite{Stern10} \\
7. & 0.200 & 72.90 & 29.60 &  \cite{Zhang14} & 26. & 0.4497 & 92.80 & 12.90 &  \cite{Moresco16}  & 44. & 0.900 & 117.0 & 23.00 &  \cite{Simon05} \\ 
8. & 0.240 & 79.69 & 3.32 &  \cite{Gaztanaga09} & 27. & 0.470 & 89.00 & 34.00 & \cite{Ratsimbazafy17} & 45. & 1.037 & 154.0 & 20.00 &  \cite{Moresco12}\\
9. & 0.270 & 77.00 & 14.00 &  \cite{Simon05}  & 28. & 0.4783 & 80.90 & 9.00 &  \cite{Moresco16} & 46. & 1.300 & 168.0 & 17.00 &  \cite{Simon05}\\
10. & 0.280 & 88.80 & 36.60 & \cite{Zhang14} & 29. & 0.480 & 87.79 & 2.03 &  \cite{Wang17} & 47. & 1.363 & 160.0 & 33.60 & \cite{Moresco15}\\ 
11. & 0.300 & 81.70 & 5.00 &  \cite{Oka14}  & 30. & 0.480 & 97.00 & 62.00 &  \cite{Stern10}  & 48. & 1.430 & 177.0 & 18.00 &  \cite{Simon05}  \\
12. & 0.310 & 78.18 & 4.74 &  \cite{Wang17} & 31. & 0.510 & 90.40 & 1.90 &  \cite{Alam17} & 49. & 1.530 & 140.0 & 14.00 &  \cite{Simon05}\\
13. & 0.340 & 83.80 & 2.96 &  \cite{Gaztanaga09} & 32. & 0.520 & 94.35 & 2.64 &  \cite{Wang17} & 50. & 1.750 & 202.0 & 40.00 &  \cite{Simon05}\\
14. & 0.350 & 82.70 & 9.10 &  \cite{Chuang13} & 33. & 0.560 & 93.34 & 2.30 &   \cite{Wang17} & 51. & 1.965 & 186.5 & 50.40 & \cite{Moresco15} \\
15. & 0.352 & 83.00 & 14.00 &  \cite{Moresco12} & 34. & 0.590 & 98.48 & 3.18 &  \cite{Wang17}  & 52. & 2.300 & 224.0 & 8.60 &  \cite{Busca13}  \\
16. & 0.360 & 79.94 & 3.38 &  \cite{Wang17} & 35. & 0.593 & 104.0 & 13.00 & \cite{Moresco12}  & 53. & 2.330 & 224.0 & 8.00 & \cite{Bautista17}\\
17. & 0.380 & 81.50 & 1.90 &  \cite{Alam17} & 36. & 0.600 & 87.90 & 6.10 &  \cite{Blake12} & 54. & 2.340 & 222.0 & 7.00 &  \cite{Delubac15} \\
18. & 0.3802 & 83.00 & 13.50 &  \cite{Moresco16} & 37. & 0.610 & 97.30 & 2.10 &  \cite{Alam17}  & 55. & 2.360 & 226.0 & 8.00 &  \cite{FontRibera14}  \\
19. & 0.400 & 95.00 & 17.00 &  \cite{Simon05} & & & & & & &&&&\\[1ex] 
\hline 
\end{tabular}
\label{table: Table II} 
\end{table*}

\section{Concluding Remarks} \label{SEC VI}
In this study, a class of modified $F(T, T_{\mathcal{G}})$ gravity models have been presented with the cosmological data sets. We first described the fundamental features of a gravitational action using a generic combination of the torsion scalar and the Gauss-Bonnet invariant. The chosen function, $F(T, T_\mathcal{G}) = -T + \lambda_1 \sqrt{T^2 + \lambda_2 T_\mathcal{G}}$, simplifies to GR as the real constant $\lambda_2$ approaches to zero. The model is based on the well-motivated Hubble parameter in the $F(T, T_{\mathcal{G}})$ gravity framework. Using the parametrization method, we also discussed the null, strong, weak, and dominant energy conditions for $F(T, T_\mathcal{G})$ gravity models. The formulae coefficients for the Hubble parameter were constrained using the Hubble dataset and the largest Pantheon SNe I dataset. It is commonly recognized that energy conditions are the best way to evaluate the self consistency of the cosmological models. We can determine whether a novel cosmological model complies with the space-time casual and geodesic structure due to the physical motivation for testing its energy conditions. We outline the major points of the current work here. Following the testing of our cosmological solutions in Sec. \ref{SEC III}, Table \ref{TABLE I} displays the values for the model parameters that best fit the data. According to the constrained values, the deceleration parameter $q$ demonstrates that the Universe makes a smooth transition from a decelerated phase of expansion to an accelerated phase in our derived model, with a deceleration-acceleration redshift of $z_{da} = 0.87$, $z_{da} = 0.77$ for Hubble and Pantheon data respectively. On the other hand, the EoS parameter indicates that the expansion of the Universe is accelerating since it is in the quintessence region. For the Hubble data and Pantheon samples, we obtained the value of the EoS parameter at $z=0$ is $\omega_0=-0.77$ and $\omega_0=-0.755$, respectively. According to the determined values of cosmological parameters and behavior, the model addressed here is more stable with the Hubble and Pantheon data set and is a feasible method for understanding the late-time acceleration of the Universe in $F(T, T_\mathcal{G})$ gravity. The extracted value of the deceleration-acceleration redshift is consistent with the few current constraints. We examined specific physical properties of the model as well as the evolution of physical parameters in combination with the energy conditions. It is observed that NEC and DEC do not violate the model, but SEC fails to fulfil it, producing a repulsive force and leading the Universe to jerk. As noted in \cite{Visser97}, the SEC violation in Fig. \ref{FIG6} and \ref{FIG7} demonstrates the viability of our model. The $Om(z)$ parameter reconstruction for the $F(T, T_\mathcal{G})$ model demonstrates that it varies between positive prior values and high positive values at the present time. The $Om(z)$ behavior suggests that our model may favor a quintessence-like behavior. Additionally, we looked at the behavior of time with redshift, which is depicted in Fig. \ref{FIG9}. It is discovered that $H_0(t_0-t)$ converges to $1.01689$ and $0.9870$ for Hubble and Pantheon data, respectively, for infinitely large $z$. This allows us to determine the age of the Universe at the present time as, $t_0 = 1.01689 H_0^{-1} \approx 14.01$ Gyrs and $t_0 = 0.9870 H_0^{-1} \approx 13.607$ Gyrs, which has been remarkably comparable to the age calculated using the Planck finding $t_0 = 13.786 \pm 0.020$ Gyrs. As a result, the model demonstrates the consistency of the accelerating evolutionary behavior of the Universe with the available data sets.
Finally we wish to mention here that since $H(z)$ and Pantheon SNe Ia are background probes, our analysis on the model applies only at the background level. In order to further distinguish this model from other, one can perform the full perturbation analysis.

\section*{Acknowledgement} SVL acknowledges the financial support provided by University Grants Commission (UGC) through Junior Research Fellowship (UGC Ref. No.:191620116597) to carry out the research work. BM acknowledges the support of Inter University Centre for Astronomy and Astrophysics (IUCAA), Pune (India) through the visiting associateship program. SKM acknowledges that this work is carried out under The Research Council (TRC) Project (Grant No. BFP/RGP/CBS-/19/099), the Sultanate of Oman. SKM is thankful for continuous support and encouragement from the administration of University of Nizwa. The authors are thankful to the honourable referee for the comments and suggestions to improve the quality of the paper.\\

\begin{widetext}
\center \appendix{\bf{Appendix}} \label{Appendix}
\begin{eqnarray}
&&\hspace{-0.3cm} \dot{H}=-\frac{1}{2} H_0^2 (z+1) \left(\beta+z (2 \zeta+3 \Omega_{m0} z + 6 \Omega_{m0})+2 \zeta+3 \Omega_{m0}\right), \nonumber\\
&&\hspace{-0.3cm} \ddot{H}= \frac{1}{2} H_0^3 (z+1) \left(\beta+z (4 \zeta+9 \Omega_{m0} z+18 \Omega_{m0})+4 \zeta+9 \Omega_{m0}\right) \sqrt{\alpha+\beta (z+1)+\zeta (z+1)^2+\Omega_{m0}(z+1)^3},\nonumber\\
&&\hspace{-0.3cm} \dot{\ddot{H}}=-\frac{1}{4} H_0^4 (z+1) \Bigg[z \Big(z \Big\{6 \Omega_{m0} (9 \alpha+34 \beta+100 \zeta)+z \Big[68 \beta \Omega_{m0}+24 \zeta^2+\Omega_{m0} z (100 \zeta+81 \Omega_{m0} z+405 \Omega_{m0}) \nonumber \\&&
  \hspace{0.3cm} +400 \zeta \Omega_{m0} +810 \Omega_{m0}^2 \Big]+24 \zeta (\beta+3 \zeta)+810 \Omega_{m0}^2 \Big\}+4 \Omega_{m0} (27 \alpha+51 \beta+100 \zeta)+16 \alpha \zeta+3 \beta^2 +48 \beta \zeta \nonumber\\&&
  \hspace{0.3cm} +72 \zeta^2 +405 \Omega_{m0}^2 \Big)+2 \alpha (\beta+8 \zeta+27 \Omega_{m0})+3 \beta^2+24 \beta \zeta+68 \beta \Omega_{m0} + 24 \zeta^2+100 \zeta \Omega_{m0}+81 \Omega_{m0}^2 \Bigg].\nonumber
\end{eqnarray}
\end{widetext}


\begin{thebibliography}{99}
\section*{References}
\bibitem{Copeland06} E. J. Copeland, M. Sami, S. Tsujikawa, \href{https://doi.org/10.1142/S021827180600942X}{Int. J. Mod. Phys. D \textbf{15}, 1753 (2006)}.

\bibitem{Cai10} Y.-F. Cai, E. N. Saridakis, M. R. Setare, J.-Q. Xia, \href{https://doi.org/10.1016/j.physrep.2010.04.001}{Phys. Rep. \textbf{493}, 1 (2010)}.

\bibitem{Capozziello11} S. Capozziello, M. De Laurentis, \href{https://doi.org/10.1016/j.physrep.2011.09.003}{Phys. Rep. \textbf{509}, 167 (2011)}.

\bibitem{Einstein28a} A. Einstein, Sitzungsber. Preuss. Akad. Wiss. Phys. Math., Kl. 217 (1928).

\bibitem{Einstein28b} A. Einstein, Sitzungsber. Preuss. Akad. Wiss. Phys. Math., Kl. 224 (1928).

\bibitem{Arcos04} H. I. Arcos, J. G. Pereira, \href{https://doi.org/10.1142/S0218271804006462}{Int. J. Mod. Phys. D \textbf{13}, 2193 (2004)}.

\bibitem{Maluf13} J. W. Maluf, \href{ https://doi.org/10.1002/andp.201200272}{Ann. Phys. \textbf{525}, 339 (2013)}.

\bibitem{Aldrovandi13} R. Aldrovandi, J. G. Pereira,\href{https://link.springer.com/book/10.1007/978-94-007-5143-9}{\textit{ Teleparallel Gravity: An Introduction, Springer, Dordrecht,} (2013)}.

\bibitem{Ferraro07} R. Ferraro, F. Fiorini, \href{https://doi.org/10.1103/PhysRevD.75.084031} {Phys. Rev. D \textbf{75}, 084031 (2007)}.

\bibitem{Bengochea09} G. R. Bengochea, R. Ferraro, \href{https://doi.org/10.1103/PhysRevD.79.124019}{Phys. Rev. D \textbf{79}, 124019 (2009)}.

\bibitem{Linder10} E. V. Linder, \href{https://doi.org/10.1103/PhysRevD.81.127301}{Phys. Rev. D \textbf{81}, 127301 (2010)}.

\bibitem{Felice10} A. De Felice, S. Tsujikawa, \href{https://doi.org/10.12942/lrr-2010-3}{Living Rev. Relativity \textbf{ 13}, 3 (2010)}.

\bibitem{Nojiri11} S. Nojiri, S. D. Odintsov, \href{https://doi.org/10.1016/j.physrep.2011.04.001}{Phys. Rep. \textbf{505}, 59 (2011)}.

\bibitem{Duchaniya22} L.K. Duchaniya, S.V. Lohakare, B. Mishra, S.K. Tripathy, \href{https://doi.org/10.1140/epjc/s10052-022-10406-w}{Eur. Phys. J. C \textbf{82}, 448 (2022)}.

\bibitem{Capozziello11a} S. Capozziello, V.F. Cardone, H. Farajollahi, and A. Ravanpak, \href{https://doi.org/10.1103/PhysRevD.84.043527}{Phys. Rev. D \textbf{84}, 043527 (2011)}.

\bibitem{Baojiu11} B. Li, T. P. Sotiriou and J. D. Barrow, \href{https://doi.org/10.1103/PhysRevD.83.064035}{Phys. Rev. D \textbf{83}, 064035 (2011)}.

\bibitem{Myrzakulov11} R. Myrzakulov, \href{https://doi.org/10.1140/epjc/s10052-011-1752-9}{Eur. Phys. J. C \textbf{71}, 1752 (2011)}.

\bibitem{Miao11} M. Li, R. X. Miao, Y.G. Miao, \href{https://doi.org/10.1007/JHEP07(2011)108}{J. High Energy Phys. \textbf{07}, 108 (2011)}.

\bibitem{Bamba11} K. Bamba, C. Q. Geng, C. C. Lee, L. W. Luo, \href{https://doi.org/10.1088/1475-7516/2011/01/021}{J. High Energy Phys. \textbf{01}, 021 (2011)}.

\bibitem{Tamanini12} N. Tamanini, C. G. B$\ddot{o}$hmer, \href{https://doi.org/10.1103/PhysRevD.86.044009}{Phy. Rev. D \textbf{86}, 044009 (2012)}.

\bibitem{Cai16} Y. F. Cai, S. Capozziello, M. De Laurentis, E. N. Saridakis, \href{https://doi.org/10.1088/0034-4885/79/10/106901}{Rep. Prog. Phys. \textbf{79}, 106901 (2016)}.

\bibitem{Anagnostopoulos19} F. K. Anagnostopoulos, S. Basilakos, E. N. Saridakis, \href{https://doi.org/10.1103/PhysRevD.103.104013}{Phys. Rev. D \textbf{100}, 083517 (2019)}.

\bibitem{DeBenedictis22} A. DeBenedictis, S. Ilijic, M. Sossich, \href{https://doi.org/10.1103/PhysRevD.105.084020}{Phys. Rev. D \textbf{105}, 084020 (2022)}.

\bibitem{Nair22} K. K. Nair, M. T. Arun, \href{https://doi.org/10.1103/PhysRevD.105.103505}{Phys. Rev. D \textbf{105}, 103505 (2022)}.


\bibitem{Riess98} A. G. Riess et al., \href{https://doi.org/10.1086/300499}{Astron. J. \textbf{116}, 1009 (1998).}

\bibitem{Perlmutter99} S. Perlmutter et al.,  \href{https://doi.org/10.1086/307221}{Astrophys. J. \textbf{517}, 565 (1999).}

\bibitem{Chevallier01} M. Chevallier, D. Polarski, \href{https://doi.org/10.1142/S0218271801000822}{Int. J. Mod. Phys. D \textbf{10}, 213 (2001)}.
.
\bibitem{Linder03} E. V. Linder, \href{https://doi.org/10.1103/PhysRevLett.90.091301}{Phys. Rev. Lett. \textbf{90}, 091301 (2003)}.

\bibitem{Lue06} A. Lue, \href{https://doi.org/10.1016/j.physrep.2005.10.007}{Phys. Rep. \textbf{423}, 1 (2006)}.

\bibitem{Nojiri07a} S. Nojiri, S. D. Odintsov, \href{https://doi.org/10.1142/S0219887807001928}{Int. J. Geom. Meth. Mod. Phys. \textbf{4}, 115 (2007)}.

\bibitem{Boulware85} D. Boulware, S. Deser, \href{https://journals.aps.org/prl/pdf/10.1103/PhysRevLett.55.2656} {Phys. Rev. Lett. \textbf{55}, 2656 (1985)}.

\bibitem{Wheeler86} J. T. Wheeler, \href{https://doi.org/10.1016/0550-3213(86)90268-3}{Nucl. Phys. B \textbf{268}, 737 (1986)}.

\bibitem{Antoniadis94} I. Antoniadis, J. Rizos, K. Tamvakis, \href{https://doi.org/10.1016/0550-3213(94)90120-1}{Nucl. Phys. B \textbf{415}, 497 (1994)}.

\bibitem{Nojiri05a} S. Nojiri, S. D. Odintsov, M. Sasaki, \href{https://doi.org/10.1103/PhysRevD.71.123509}{Phys. Rev. D \textbf{71}, 123509 (2005)}.
  
\bibitem{Kofinas14} G. Kofinas, E. N. Saridakis, \href{https://doi.org/10.1103/PhysRevD.90.084044}{Phys. Rev. D \textbf{90}, 084044 (2014)}.

\bibitem{Kofinas14a} G. Kofinas, E.N. Saridakis, \href{https://doi.org/10.1103/PhysRevD.90.084045}{Phys. Rev. D \textbf{90}, 084045 (2014).}

\bibitem{Harko14a} T. Harko, F. S. N. Lobo, G. Otalora and E. N. Saridakis, \href{https://doi.org/10.1088/1475-7516/2014/12/021}{J. Cosmol. Astropart. Phys. \textbf{12}, 021 (2014)}.

\bibitem{Harko14b} T. Harko, F. S. N. Lobo, G. Otalora and E. N. Saridakis, \href{https://doi.org/10.1103/PhysRevD.89.124036}{Phys. Rev. D \textbf{89}, 124036 (2014)}.

\bibitem{Kofinas14b} G. Kofinas, G. Leon and
E.N. Saridakis, \href{https://doi.org/10.1088/0264-9381/31/17/175011}{Class. Quantum Grav. \textbf{31}, 175011 (2014).}

\bibitem{Chattopadhyay14} S. Chattopadhyay et al., \href{https://doi.org/10.1007/s10509-014-2029-1}{Astrophys. \& Space Sci. \textbf{353}, 279 (2014).}

\bibitem{Capozziello16} S. Capozziello, M. De Laurentis and K. F. Dialektopoulos, \href{https://doi.org/10.1140/epjc/s10052-016-4491-0}{Eur. Phys. J. C \textbf{76}, 629 (2016)}.

\bibitem{Foreman13} D. Foreman-Mackey et al., \href{https://doi.org/10.1086/670067}{Publ. Astron. Soc. Pac. \textbf{125}, 306 (2013).}

\bibitem{Scolnic18} D.M. Scolnic et al., \href{https://ui.adsabs.harvard.edu/link_gateway/2018ApJ...859..101S/doi:10.3847/1538-4357/aab9bb}{Astrophys. J. \textbf{859}, 101 (2018).}

\bibitem{Yu18} H. Yu, B. Ratra and F.-Y. Wang, \href{https://doi.org/10.3847/1538-4357/aab0a2}{Astrophys. J. \textbf{856}, 3 (2018).}

\bibitem{Sahni03} V. Sahni, T.D. Saini, A.A. Starobinsky and U. Alam, \href{https://doi.org/10.1134/1.1574831}{J. Exp. \& Theor. Phys. Lett. \textbf{77}, 201 (2003).}

\bibitem{Capozziello05} S. Capozziello, V.F. Cardone, A. Troisi, \href{https://doi.org/10.1103/PhysRevD.71.043503}{Phys. Rev. D \textbf{71}, 043503 (2005).}

\bibitem{Saini00} T.D. Saini, S. Raychaudhury, V. Sahni, A.A. Starobinsky, \href{https://doi.org/10.1103/PhysRevLett.85.1162}{Phys. Rev. Lett. \textbf{85}, 1162 (2000).}

\bibitem{Sharov18} G.S. Sharov, V.O. Vasiliev, \href{http://mmg.tversu.ru/images/publications/2018-611}{Math. Model. Geom. \textbf{6}, 1 (2018).}

\bibitem{Ferreira97} P.G. Ferreira and M. Joyce, \href{https://doi.org/10.1103/PhysRevLett.79.4740}{Phys. Rev. Lett. \textbf{79}, 4740 (1997).}

\bibitem{Copeland98} E.J. Copeland, A.R. Liddle and D. Wands, \href{https://doi.org/10.1103/PhysRevD.57.4686}{Phys. Rev. D \textbf{57}, 4686 (1998).}

\bibitem{Chen09} Xi-m. Chen, Y. Gong and E. N. Saridakis, \href{https://doi.org/10.1088/1475-7516/2009/04/001}{J. Cosmol. Astropart. Phys. \textbf{04}, 001 (2009).}

\bibitem{Gruber14} Christine Gruber and Orlando Luongo, \href{http://dx.doi.org/10.1103/PhysRevD.89.103506}{Phys. Rev. D \textbf{89}, 103506 (2014).}

\bibitem{Busca13} N.G. Busca et al., \href{https://doi.org/10.1051/0004-6361/201220724}{Astron. \& Astrophys. \textbf{552}, A96 (2013).}

\bibitem{Farooq13} O. Farooq, B. Ratra, \href{http://dx.doi.org/10.1088/2041-8205/766/1/L7}{Astrophys. J. Lett. \textbf{766}, L7 (2013).}

\bibitem{Lu11} J. Lu, L. Xu, M. Liu, \href{https://doi.org/10.1016/j.physletb.2011.04.022}{Phys. Lett. B \textbf{699}, 246 (2011).}

\bibitem{Capozziello14} S. Capozziello, O. Farooq, O. Luongo, B. Ratra, \href{https://doi.org/10.1103/PhysRevD.90.044016}{Phys. Rev. D \textbf{90}, 044016 (2014).}

\bibitem{Yang20} Y. Yang and Y. Gong, \href{https://doi.org/10.1088/1475-7516/2020/06/059} {J. Cosmol. Astropart. Phys. \bf{06}, 059 (2020).}

\bibitem{Amanullah10} Amanullah et al., \href{http://dx.doi.org/10.1088/0004-637X/716/1/712}{Astrophys. J. Lett. \textbf{716}, 712 (2010).}

\bibitem{Valentino16} E. Di Valentino, A. Melchiorri, and J. Silk, \href{https://doi.org/10.1016/j.physletb.2016.08.043}{Phys. Lett. B \textbf{761}, 242 (2016)}.

\bibitem{Vagnozzi20} S. Vagnozzi, \href{https://journals.aps.org/prd/abstract/10.1103/PhysRevD.102.023518}{Phys. Rev. D \textbf{102}, 023518 (2020)}.

\bibitem{Efstathiou20} G. Efstathiou and S. Gratton, \href{https://doi.org/10.1093/mnrasl/slaa093}{Month. Not. Roy. Astron. Soc.: Lett. \textbf{496}, L91 (2020)}.

\bibitem{Vagnozzi21} S. Vagnozzi, A. Loeb, M. Moresco, \href{https://iopscience.iop.org/article/10.3847/1538-4357/abd4df}{Astrophys. J. \textbf{908}, 84 (2021)}.

\bibitem{Santos07} J. Santos et al., \href{https://doi.org/10.1103/PhysRevD.76.083513}{Phys. Rev. D \textbf{76}, 083513 (2007).}

\bibitem{Kar07} S. Kar and S. SenGupta, \href{https://doi.org/10.1007/s12043-007-0110-9}{Pramana \textbf{69}, 49 (2007).}

\bibitem{Hawking73}  S. Hawking and G.F.R. Ellis, \href{https://books.google.co.in/books?hl=en&lr=&id=QagG_KI7Ll8C&oi=fnd&pg=PP15&dq=S.+Hawking+and+G.F.R.+Ellis,+The+Large+Scale+Structure+of+Space-Time,+Cambridge+University+Press,+(1973).&ots=GqnlnlPATa&sig=kehrRusxQAZgl8ZruhZjR8Gs1Wg&redir_esc=y#v=onepage&q=S.\%20Hawking\%20and\%20G.F.R.\%20Ellis\%2C\%20The\%20Large\%20Scale\%20Structure\%20of\%20Space-Time\%2C\%20Cambridge\%20University\%20Press\%2C\%20(1973).&f=false} {\textit{The Large Scale Structure of Space-Time, Cambridge University Press,} (1973).}

\bibitem{Poisson04}  E. Poisson, \href{https://books.google.co.in/books?hl=en&lr=&id=bk2XEgz_ML4C&oi=fnd&pg=PP1&dq=E.+Poisson,+A+Relativist\%E2\%80\%99s+Toolkit:+The+Mathematics+of+Black+Hole+Mechanics,+Cambridge+University+Press,+(2004).&ots=d65cN_k24H&sig=SZKENebqgib_vJ7Tg_9C_XRlUMg&redir_esc=y#v=onepage&q=E.\%20Poisson\%2C\%20A\%20Relativist\%E2\%80\%99s\%20Toolkit\%3A\%20The\%20Mathematics\%20of\%20Black\%20Hole\%20Mechanics\%2C\%20Cambridge\%20University\%20Press\%2C\%20(2004).&f=false} {\textit{A Relativist's Toolkit: The Mathematics of Black Hole Mechanics, Cambridge University Press,} (2004).}

\bibitem{Barcelo02} C. Barcelo and M. Visser, \href{https://doi.org/10.1142/S0218271802002888}{Int. J. Mod. Phys. D \textbf{11}, 1553 (2002).} 

\bibitem{Visser97} M. Visser, \href{https://doi.org/10.1103/PhysRevD.56.7578}{Phys. Rev. D \textbf{56}, 7578 (1997).}

\bibitem{Sahni08} V. Sahni, A. Shafieloo, A.A. Starobinsky, \href{https://doi.org/10.1103/PhysRevD.78.103502}{Phys. Rev. D \textbf{78}, 103502 (2008).}

\bibitem{Sahni14} V. Sahni, A. Shafieloo, A.A. Starobinsky, \href{https://doi.org/10.1088/2041-8205/793/2/L40}{Astrophys. J. \textbf{L40}, 793 (2014).}

\bibitem{Ding15} X. Ding et al., \href{https://doi.org/10.1088/2041-8205/803/2/L22}{Astrophys. J. Lett. \textbf{803}, L22 (2015).}

\bibitem{Zheng16} X. Zheng et al., \href{https://doi.org/10.3847/0004-637X/825/1/17}{Astrophys. J. \textbf{825}, 17 (2016).}

\bibitem{Qi18} Jing-Zhao Qi et al., \href{https://doi.org/10.1088/1674-4527/18/6/66}{Res. Astron. \& Astrophys. \textbf{18}, 066 (2018).}

\bibitem{Gribbin15} Gribbin, J., \href{https://www.perlego.com/book/569711/138-the-quest-to-find-the-true-age-of-the-universe-and-the-theory-of-everything-pdf}{\textit{13.8, the quest to find the true age of the Universe and the theory of everything. London: Icon Books Ltd.} (2015).}

\bibitem{Aghanim20} N. Aghanim Planck Collaboration et al., \href{https://doi.org/10.1051/0004-6361/201833910}{Astron. \& Astrophys. \textbf{641} A6 (2020).}

 \bibitem{Vagnozzi22} S. Vagnozzi, F. Pacucci, A. Loeb, \href{https://doi.org/10.1016/j.jheap.2022.07.004}{J. High Energy Astrophys. \textbf{36}, 27 (2022)}.

\bibitem{Zhang14} C. Zhang et al., \href{https://doi.org/10.1088/1674-4527/14/10/002}{Res. Astron. \& Astrophys. \textbf{14}, 1221 (2014).}

\bibitem{Jimenez03} R. Jimenez, L. Verde, T. Treu, and D. Stern, \href{https://doi.org/10.1086/376595}{Astrophys. J. \textbf{593}, 622 (2003).}

\bibitem{Simon05} J. Simon, L. Verde and R. Jimenez, \href{https://doi.org/10.1103/PhysRevD.71.123001}{Phys. Rev. D \textbf{71}, 123001 (2005).}

\bibitem{Moresco12} M. Moresco et al. \href{https://doi.org/10.1088/1475-7516/2012/08/006}{J. Cosmol. Astropart. Phys. \textbf{08}, 006 (2012).}

\bibitem{Gaztanaga09} E. Gaztanaga et al. \href{https://doi.org/10.1111/j.1365-2966.2009.15405.x}{Mon. Not. R. Astron. Soc. \textbf{399}, 1663 (2009).}

\bibitem{Oka14} A. Oka et al., \href{https://doi.org/10.1093/mnras/stu111}{Mon. Not. R. Astron. Soc. \textbf{439}  2515 (2014).}

\bibitem{Wang17} Y. Wang et al., \href{https://doi.org/10.1093/mnras/stx1090}{Mon. Not. R. Astron. Soc. \textbf{469}, 3762 (2017).}

\bibitem{Chuang13} C. H. Chuang and Y. Wang, \href{https://doi.org/10.1093/mnras/stt1290}{Mon. Not. R. Astron. Soc. \textbf{435}, 255 (2013).}

\bibitem{Alam17} S. Alam et al., \href{https://doi.org/10.1093/mnras/stx721}{Mon. Not. R. Astron. Soc. \textbf{470}, 2617 (2017).}

\bibitem{Moresco16} M. Moresco et al., \href{https://doi.org/10.1088/1475-7516/2016/05/014}{J. Cosmol. Astropart. Phys. \textbf{05}, 014 (2016).}

\bibitem{Blake12} C. Blake et al., \href{https://doi.org/10.1111/j.1365-2966.2012.21473.x}{Mon. Not. R. Astron. Soc. \textbf{425}, 405 (2012).}

\bibitem{Ratsimbazafy17} A.L. Ratsimbazafy et al., \href{https://doi.org/10.1093/mnras/stx301}{Mon. Not. R. Astron. Soc. \textbf{467}, 3239 (2017).}

\bibitem{Stern10} D. Stern et al., \href{https://doi.org/10.1088/1475-7516/2010/02/008}{J. Cosmol. Astropart. Phys. \textbf{02}, 008 (2010).}

\bibitem{Moresco15} M. Moresco, \href{https://doi.org/10.1093/mnrasl/slv037}{Mon. Not. R. Astron. Soc. \textbf{450}, L16 (2015).}

\bibitem{Bautista17} J.E. Bautista et al., \href{https://doi.org/10.1051/0004-6361/201730533}{Astron. \& Astrophys. \textbf{603}, A12 (2017).}

\bibitem{Delubac15} T. Delubac et al., \href{https://doi.org/10.1051/0004-6361/201423969}{Astron. \& Astrophys. \textbf{574}, A59 (2015).}

\bibitem{FontRibera14} A. Font-Ribera et al., \href{https://doi.org/10.1088/1475-7516/2014/05/027}{J. Cosmol. Astropart. Phys. \textbf{05}, 027 (2014).}

\end{thebibliography}
\end{document}